\documentclass[twocolumn,preprintnumbers,superscriptaddress,amsmath,amssymb,nofootinbib]{revtex4}

\usepackage{graphicx}
\usepackage{dcolumn}
\usepackage{bm}
\usepackage{amsmath, amssymb}

\allowdisplaybreaks[1]

\usepackage[normalem]{ulem}  

\renewcommand\sout{\bgroup \color{red} \ULdepth=-.5ex \ULset}

\newcommand{\Slash}[1]{\ooalign{\hfil/\hfil\crcr$#1$}}
\newcommand{\Psfig}[2]{\includegraphics[width=#1]{#2}}
\newcommand{\PsfigII}[2]{\includegraphics[scale=#1]{#2}}

\newcommand{\SUN}[1]{\text{SU} ( #1 )}

\def\Schr{Schr\"{o}dinger }

\def\mev{\text{~MeV}}
\def\gev{\text{~GeV}}
\def\fm{\text{~fm}}

\newcommand{\wb}{\rm wb}

\begin{document}

\preprint{YITP-18-39}

\title{\boldmath $N \Omega$ interaction: meson exchanges, inelastic
  channels, and quasibound state}

\author{Takayasu Sekihara} 
\email{sekihara@post.j-parc.jp}
\affiliation{Advanced Science Research Center, Japan Atomic Energy
  Agency, Shirakata, Tokai, Ibaraki, 319-1195, Japan}

\author{Yuki Kamiya}
\email{yuki.kamiya@yukawa.kyoto-u.ac.jp}
\affiliation{Yukawa Institute for Theoretical Physics, Kyoto
  University, Kyoto 606-8502, Japan}

\author{Tetsuo Hyodo}
\email{hyodo@yukawa.kyoto-u.ac.jp}
\affiliation{Yukawa Institute for Theoretical Physics, Kyoto
  University, Kyoto 606-8502, Japan}

\date{\today}

\begin{abstract}

  Based on a baryon--baryon interaction model with meson exchanges, we
  investigate the origin of the strong attraction in the $N \Omega (
  {}^{5}S_{2} )$ interaction, which was indicated by recent lattice
  QCD simulations.  The long range part of the potential is
  constructed by the conventional mechanisms, the exchanges of the
  $\eta$ meson and of the correlated two mesons in the
  scalar-isoscalar channel, denoted by ``$\sigma$'' in the literature.
  The short range part is represented by the contact interaction.  We
  find that the meson exchanges do not provide sufficient
  attraction. This means that most of the attraction is attributed to
  the short range contact interaction.  We then evaluate the effect of
  the coupled channels to the $N \Omega ( {}^{5}S_{2} )$ interaction.
  We find that, while the $D$-wave mixing of the $N \Omega$ channel is
  negligible, the inelastic $\Lambda \Xi$, $\Sigma \Xi$, and $\Lambda
  \Xi (1530)$ channels via the $K$ meson exchange give the attraction
  of the $N \Omega ( {}^{5}S_{2} )$ interaction to the same level with
  the elastic meson exchanges. Although the elimination of these
  channels induces the energy dependence of the single-channel $N
  \Omega$ interaction, this effect is not significant.  With the
  present model parameters fitted to reproduce the scattering length
  of the HAL QCD result of the nearly physical quark masses, we obtain
  the $N \Omega ( {}^{5}S_{2} )$ quasibound state with its eigenenergy
  $2611.3 - 0.7 i \mev$, which corresponds to the binding energy $0.1
  \mev$ and width $1.5 \mev$ for the decay to the $\Lambda \Xi$ and
  $\Sigma \Xi$ channels.  From the analysis of the spatial structure
  and the compositeness, the quasibound state is shown to be the
  molecular state of $N\Omega$. We also construct an equivalent local
  potential for the $N \Omega ( {}^{5}S_{2} )$ system which is useful
  for various applications.

\end{abstract}

\maketitle

\section{Introduction}

Existence and properties of dibaryons have been one of the major
topics in hadron physics.  Here dibaryons stand for states of baryon
number $B=2$ generated by strong interactions regardless of their
structure: compact hexaquarks, baryon--baryon and
meson--baryon--baryon molecules, etc.  So far, there is only single
well-established dibaryon state, the deuteron, which is a
proton--neutron molecule~\cite{Weinberg:1965zz}.  Because various
different mechanisms in strong interactions are expected to generate
dibaryons, the study of dibaryons helps to
understand the underlying theory of strong interactions, quantum
chromodynamics (QCD). For instance, compact hexaquarks are closely
related to the mechanism of quark confinement and correlation of
quarks inside hadrons.  Hadronic molecules serve as a valuable clue to
investigate the hadron--hadron interactions which lead to novel
few-body systems bound by hadronic interactions.

Historically, dibaryons were first discussed in theoretical studies.
In the early stage, dibaryons analogous to the deuteron were predicted
in Ref.~\cite{Dyson:1964xwa} by combining the nucleon ($N$) and the
$\Delta$ resonance.  The $H$ dibaryon was predicted as a compact
hexaquark owing to the strongly attractive color-magnetic interaction
between quarks~\cite{Jaffe:1976yi}, which stimulated various
experimental searches for such dibaryons (see
review~\cite{Clement:2016vnl}).  As an example of the dibaryon with
meson--baryon--baryon structure, the $\bar{K}NN$ state was
predicted~\cite{Yamazaki:2002uh}, motivated by the strong attraction
between the antikaon ($\bar{K}$) and nucleon~\cite{Akaishi:2002bg}
(see reviews~\cite{Hyodo:2011ur,Gal:2016boi}).  Then, recent
remarkable progress in experiments and in lattice QCD simulations
enables us to examine these theoretical predictions on dibaryons.  For
example, the WASA-at-COSY collaboration has recently reported the
$d^{\ast} (2380)$ in quantum numbers $(J^{P} , \, I ) = ( 3^{+} , \, 0
)$~\cite{Bashkanov:2008ih, Adlarson:2011bh, Adlarson:2012fe}, which
may correspond to the isoscalar $\Delta \Delta$ bound state predicted
in Ref.~\cite{Dyson:1964xwa}.  Some hints about the $H$ dibaryon come
from lattice QCD simulations~\cite{Beane:2010hg, Inoue:2010es,
  Beane:2011iw, Inoue:2011ai, Sasaki:2013zwa, Sasaki:2015ifa,
  Sasaki:2016gpc, Sasaki:2017ysy}.  In particular, the baryon--baryon
interaction in the HAL QCD method with nearly physical quark
masses~\cite{Sasaki:2016gpc, Sasaki:2017ysy} implies existence of a
resonance around the $N\Xi$ threshold.  Lattice QCD analyses with the
nearly physical quark masses are further suggesting new dibaryons such
as the $S$-wave $\Omega \Omega$ bound state in $J^{P} =
0^{+}$~\cite{Gongyo:2017fjb}.  The J-PARC E15 experiment observed a
peak structure which can be interpreted as a signal of the $\bar{K} N
N$ quasibound state~\cite{Sada:2016nkb, Sekihara:2016vyd}.

\begin{table}[b]
  \caption{Baryon--baryon channels coupling to $N \Omega$ and their
    threshold energies. }
  \label{tab:chan}
  \begin{ruledtabular}
    \begin{tabular}{cc}
      Channel & Threshold [MeV] 
      \\
      \hline
      $\Lambda \Xi$ & 2434
      \\
      $\Sigma \Xi$ & 2511
      \\
      $N \Omega$ & 2611
      \\
      $\Lambda \; \Xi (1530) = \Lambda \Xi ^{\ast}$ & 2649
      \\
      $\Sigma (1385) \; \Xi = \Sigma ^{\ast} \Xi$ & 2703
      \\
      $\Sigma \; \Xi (1530) = \Sigma \Xi ^{\ast}$ & 2727
      \\
      $\Sigma (1385) \; \Xi (1530) = \Sigma ^{\ast} \Xi ^{\ast}$ & 2918
    \end{tabular}
  \end{ruledtabular}
\end{table}

\begin{figure*}[!t]
  \centering
  \PsfigII{0.215}{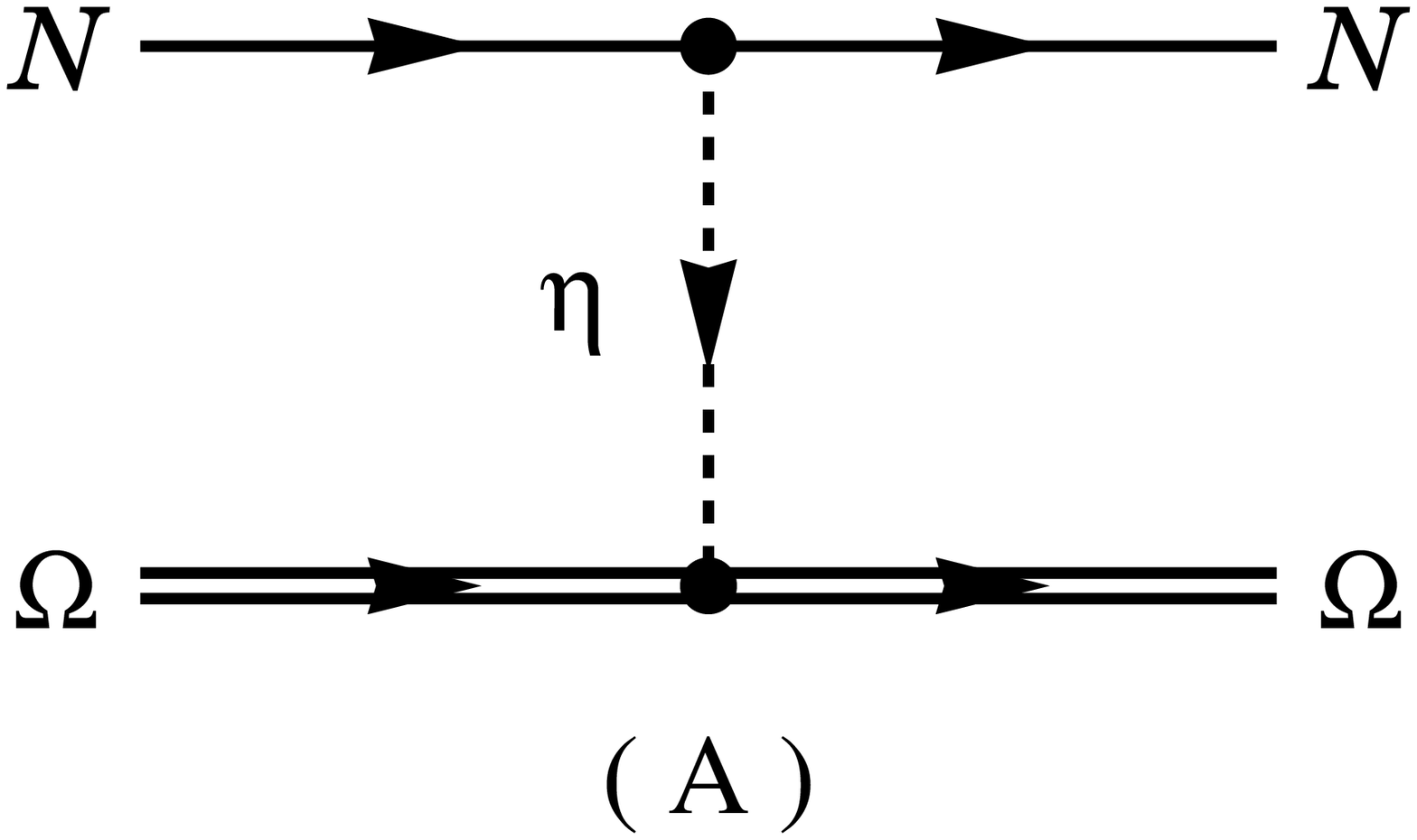} ~
  \PsfigII{0.215}{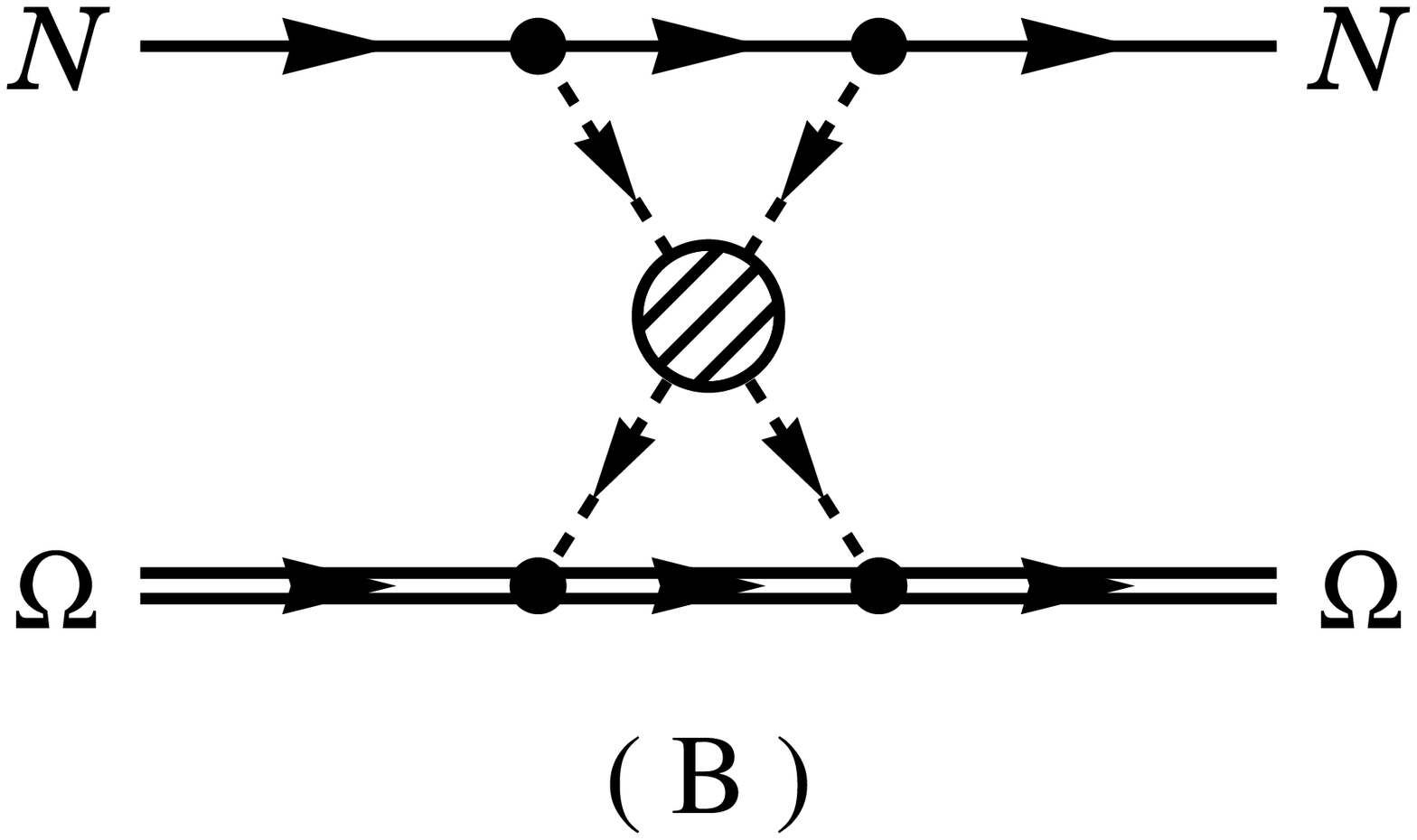} ~
  \PsfigII{0.215}{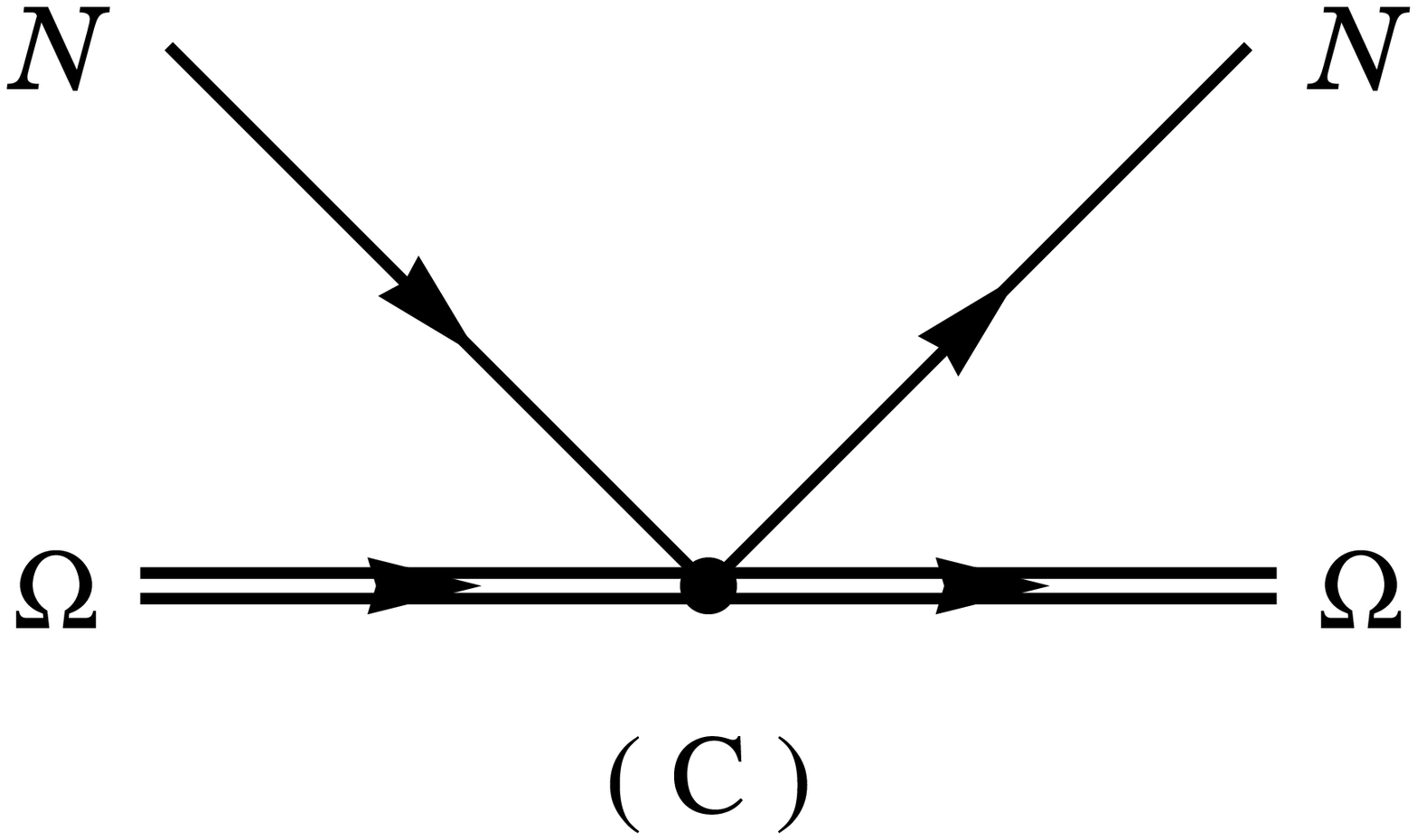} ~
  \PsfigII{0.215}{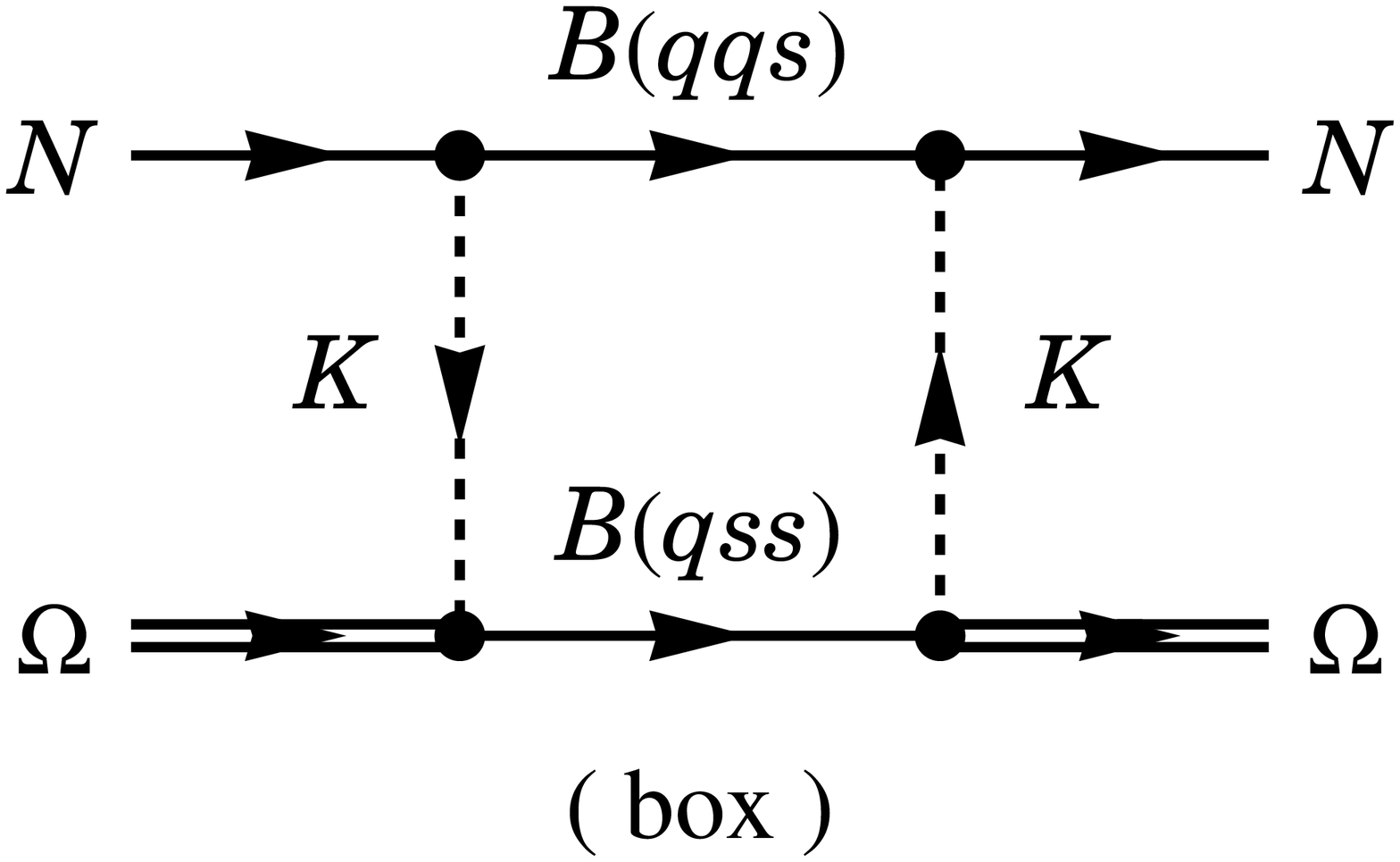}  
  \caption{Feynman diagrams for the $N \Omega$ interaction.  The
    dashed lines represent the pseudoscalar mesons, while the solid
    and double lines indicate baryons.  Shaded circle denotes the
    correlation of two mesons, and $B ( q q s ) B ( q s s )$
    represents $\Lambda \Xi$, $\Sigma \Xi$, and $\Lambda \Xi (1530)$.}
  \label{fig:V_NO}
\end{figure*}

In this study, we focus on yet another dibaryon system, the $N \Omega$
state.  This system was predicted to be bound in quark-model
calculations~\cite{Goldman:1987ma, Oka:1988yq, Li:1999bc,
  Pang:2003ty}, and further studies within quark models are found in
Refs.~\cite{Zhu:2015sna, Huang:2015yza}. Remarkably, the repulsive
core is expected to be absent in the elastic $N\Omega$ potential, in
contrast to the nuclear force, because the quark flavors in $N$ are
completely different from those in $\Omega$ and hence the Pauli
exclusion principle does not work.  The absence of the repulsive core
is advantageous to generate a possible dibaryon state in the $N
\Omega$ system.  The $N \Omega$ interaction in the ${}^{5} S_{2}$
channel was recently obtained in the HAL QCD analyses of the lattice
QCD data, where ${}^{2 S + 1} L_{J}$ denotes the state with spin $S$,
$L$ wave, and total angular momentum $J$ of the $N \Omega$ system.
Interestingly, the results of the HAL QCD analyses suggested a
strongly attractive potential in the $N \Omega ({}^{5} S_{2})$ channel
without repulsive core which supports a bound
state~\cite{Etminan:2014tya, Doi:2017zov, Iritani:2018a}.  Although
there are lower energy baryon--baryon coupled channels $\Lambda \Xi$
and $\Sigma \Xi$, it is expected that the decay of the $N \Omega
({}^{5} S_{2})$ quasibound state will be suppressed because couplings
to these decay channels are in $D$ wave (see Table~\ref{tab:chan} for
baryon--baryon channels coupling to the $N \Omega$ state).  Stimulated
by the HAL QCD results, the $N\Omega$ interaction was studied in the
framework of chiral perturbation theory~\cite{Haidenbauer:2017sws}.  A
method to probe this dibaryon with the correlation between $N$ and
$\Omega$ in high-energy heavy ion collisions was proposed in
Ref.~\cite{Morita:2016auo} as well.

The aim of our study is to understand the origin of the strong
attraction in the $N \Omega ({}^{5} S_{2})$ channel.  For this
purpose, we construct a meson exchange model for the $N \Omega$
interaction.  Combining the long-range meson exchange mechanisms with
the short-range interaction represented by the contact term, we can
pin down the physical origin of the attractive $N\Omega$
interaction. In addition, by taking into account the coupling to the
relevant baryon--baryon inelastic channels, we can further discuss the
absorption processes and the energy dependence of the $N \Omega$
interaction. These effects was assumed to be small and neglected in
the HAL QCD analyses of the $N \Omega$ interaction.  Finally, the
attractive $N\Omega$ interaction implies the possible existence of
nuclei with an $\Omega$ baryon. It is practically useful to construct
a local potential equivalent to the full model, for the application to
few-body calculations of $\Omega$ nuclei.

This paper is organized as follows.  First, in Sec.~\ref{sec:2} we
formulate the $N \Omega$ interaction including the inelastic
contributions as well as the elastic channels.  Next, we show the
expression of the scattering amplitude and determine the model
parameters so as to reproduce the $N \Omega ( {}^{5} S_{2} )$
scattering length calculated in the HAL QCD analyses in
Sec.~\ref{sec:3}.  We then discuss the $N \Omega ( {}^{5} S_{2} )$
interaction in Sec.~\ref{sec:4} by separately evaluating the elastic
and inelastic contributions to the interaction.  We also calculate
properties of the on-shell $N \Omega$ scattering amplitude and of the
$N \Omega$ quasibound state.  In Sec.~\ref{sec:5} we construct an
equivalent local potential which reproduces the $N \Omega ( {}^{5}
S_{2} )$ scattering amplitude.  Section~\ref{sec:6} is devoted to the
conclusion of this study.

\section{\boldmath Formulation of the $N \Omega$ interaction}
\label{sec:2}

First of all, we formulate the $N \Omega$ interaction based on the
meson exchanges with effective Lagrangians. This interaction is then
used to obtain the scattering amplitude in Sec.~\ref{sec:3}.

\subsection{Mechanisms}
\label{sec:2A}

As for the elastic $N\Omega$ channel, the Okubo--Zweig--Iizuka (OZI)
rule restricts mediating mesons to those containing both $( u \bar{u}
+ d \bar{d} ) / \sqrt{2}$ and $s \bar{s}$ components.  Owing to this
fact, the longest range interaction should be mediated by the $\eta$
exchange [Fig.~\ref{fig:V_NO}(A)].  In addition to $\eta$, there is a
contribution from the exchange of the light scalar-isoscalar meson
``$\sigma$'', which should be, however, treated as the exchange of
correlated two pseudoscalar mesons due to its broad width as shown in
Fig.~\ref{fig:V_NO}(B).  In the vector channel, on the other hand,
the exchange of the light vector mesons is forbidden, because of their
ideal mixing and the OZI rule.  The contributions from the $\eta$ and
correlated two-meson exchanges can be determined by empirical information as
we show below.  Further contributions at short ranges, such as the
exchanges of the heavier mesons and the color magnetic interactions at
quark-gluon level, are treated as a contact term
[Fig.~\ref{fig:V_NO}(C)].

There are several inelastic channels which can couple to $N\Omega$ as
shown in Table~\ref{tab:chan}. Among them, we take into account the
two open channels $\Lambda \Xi$ and $\Sigma \Xi$ which are responsible
for the absorption processes.  We also include one closed channel
$\Lambda \Xi ^{\ast}$, whose threshold is nearest to the $N \Omega$
threshold.  We consider the transition from $N \Omega$ to these
channels through the $K$ exchange.  We expect that, around the
$N\Omega$ threshold, the transitions between the inelastic channels
such as $\Lambda \Xi \to \Lambda \Xi$ contributes to the $N \Omega$
interaction only subdominantly.  Neglecting these contributions, we
evaluate the box diagrams to include the inelastic effects on the
$N\Omega$ interaction as shown in Fig.~\ref{fig:V_NO}(box).

\subsection{Effective Lagrangians}
\label{sec:2B}

The vertices in Figs.~\ref{fig:V_NO} are constructed with the
effective Lagrangians including the pseudoscalar meson $P$, octet
baryon $B$, and decuplet baryon $D$, based on flavor $\SUN{3}$
symmetry.

The $P B B$ coupling is governed by the chiral Lagrangian:
\begin{equation}
  \mathcal{L} = - \frac{F}{\sqrt{2} f} 
  \left \langle \bar{\mathcal{B}} \gamma ^{\mu} \gamma _{5}
       \left [ \partial _{\mu} \varPhi , \, \mathcal{B} \right ] \right \rangle
  - \frac{D}{\sqrt{2} f} 
  \left \langle \bar{\mathcal{B}} \gamma ^{\mu} \gamma _{5}
  \left \{ \partial _{\mu} \varPhi , \, \mathcal{B} \right \} \right \rangle ,
  \label{eq:chiL}
\end{equation}
with the pseudoscalar meson and octet baryon fields $\varPhi$ and
$\mathcal{B}$, respectively, whose explicit forms are
\begin{equation}
  \varPhi = \left ( 
  \begin{array}{@{\,}ccc@{\,}}
    \frac{1}{\sqrt{2}} \pi ^{0} + \frac{1}{\sqrt{6}} \eta &
    \pi^{+} & K^{+} \\
    \pi^{-} & - \frac{1}{\sqrt{2}} \pi ^{0} + \frac{1}{\sqrt{6}} \eta &
    K^{0} \\ K^{-} & \bar{K} ^{0} & - \frac{2}{\sqrt{6}} \eta 
  \end{array}
  \right ) ,
\end{equation}
\begin{equation}
\mathcal{B} = \left( 
\begin{array}{@{\,}ccc@{\,}}
\frac{1}{\sqrt{2}} \Sigma ^{0} + \frac{1}{\sqrt{6}} \Lambda  &
\Sigma ^{+} & p \\
\Sigma ^{-} & - \frac{1}{\sqrt{2}} \Sigma ^{0} + \frac{1}{\sqrt{6}}
 \Lambda &
n \\ \Xi ^{-} & \Xi ^{0} & - \frac{2}{\sqrt{6}} \Lambda  
\end{array}
\right) . 
\end{equation}
The meson decay constant $f$ is chosen at their physical
values~\cite{Olive:2016xmw}: $f_{\pi} = 92.1 \mev$, $f_{K} = 1.2
f_{\pi}$, and $f_{\eta} = 1.3 f_{\pi}$.  The parameters $D = 0.795$
and $F = 0.465$ are fixed by the weak decays of the octet baryons.

The Lagrangian for the $P B D$ coupling is
\begin{equation}
  \mathcal{L} = - \frac{f_{P B D}}{m_{\pi}} \left \langle 
  \left ( \bar{\Delta}_{\mu} \cdot \partial ^{\mu} \varPhi \right )
  \mathcal{B}
  + \text{h.c.} \right \rangle ,
\label{eq:LPBD}
\end{equation}
where $m_{\pi}$ is the pion mass and the product $( \bar{\Delta} \cdot
\varPhi )$ represents
\begin{equation}
  ( \bar{\Delta} \cdot \varPhi )_{a b} = 
  \epsilon _{i j a} \bar{\Delta}_{i k b} \varPhi _{k j} ,
\end{equation}
with the decuplet baryon field $\Delta$:
\begin{equation}
  \begin{split}
    & \Delta _{111} = \Delta ^{++} ,
    \quad 
    \Delta _{112} = \frac{1}{\sqrt{3}} \Delta ^{+} ,
    \\
    & \Delta _{122} = \frac{1}{\sqrt{3}} \Delta ^{0} ,
    \quad 
    \Delta _{222} = \Delta ^{-} ,
  \end{split}
\end{equation}
for $\Delta (1232)$, 
\begin{equation}
  \Delta _{113} = \frac{1}{\sqrt{3}} \Sigma ^{\ast +} ,
  \quad
  \Delta _{123} = \frac{1}{\sqrt{6}} \Sigma ^{\ast 0} ,
  \quad
  \Delta _{223} = \frac{1}{\sqrt{3}} \Sigma ^{\ast -} ,
\end{equation}
for $\Sigma (1385)$, 
\begin{equation}
  \Delta _{133} = \frac{1}{\sqrt{3}} \Xi ^{\ast 0} ,
  \quad
  \Delta _{233} = \frac{1}{\sqrt{3}} \Xi ^{\ast -} ,
\end{equation}
for $\Xi (1530)$, and
\begin{equation}
  \Delta _{333} = \Omega ^{-} ,
\end{equation}
for $\Omega ^{-}$.  The form of $\Delta _{a b c}$ is completely
symmetric under permutations of indices $a$, $b$, and $c = 1$, $2$,
$3$.  The coupling constant $f_{P B D}$ is fixed as $f_{P B D} = 1.8$
so as to reproduce semi-quantitatively the decay widths of the
decuplet baryons: with $f_{P B D} = 1.8$, we obtain $\Gamma _{\Delta
  (1232) \to \pi N} = 63 \mev$, $\Gamma _{\Sigma (1385) \to \pi
  \Lambda} = 33 \mev$, $\Gamma _{\Sigma (1385) \to \pi \Sigma} = 5
\mev$, and $\Gamma _{\Xi (1530) \to \pi \Xi} = 14 \mev$.

Similarly, the Lagrangian for the $P D D$ coupling is
\begin{equation}
  \mathcal{L} = - \frac{f_{P D D}}{m_{\pi}}
  \left \langle
  ( \bar{\Delta}^{\mu} \cdot \gamma ^{\nu} \gamma _{5} \Delta _{\mu} )
  \partial _{\nu} \varPhi 
  \right \rangle .
\label{eq:LPDD}
\end{equation}
where the product $( \bar{\Delta} \cdot \Delta )$ represents
\begin{equation}
  ( \bar{\Delta} \cdot \Delta )_{a b} = 
  \bar{\Delta}_{i j b} \Delta _{i j a} .
\end{equation}
The coupling constant $f_{P D D}$ is fixed from the nucleon axial
charge based on the nonrelativistic $\SUN{6}$ quark
model~\cite{Brown:1975di}:
\begin{equation}
  \frac{g_{A}}{2}
  : \frac{f \times f_{P D D}}{\sqrt{2} m_{\pi}}
  = \frac{5}{6}
  : \frac{3}{2} ,
\end{equation}
where $g_{A} \equiv D + F = 1.26$.  From this analysis, we obtain
$f_{P D D} = 9 g_{A} m_{\pi} / (5 \sqrt{2} f) = 2.09$.

Finally, we employ a spin-independent form for the contact $B D B D$
Lagrangian
\begin{equation}
  \mathcal{L} =
  c \left ( \bar{\Omega} \Omega \right )
  \left ( \bar{p} p + \bar{n} n \right ) ,
  \label{eq:Lcontact}
\end{equation}
with a coupling constant $c$ as a model parameter.  In general there
is a spin-dependent contact $B D B D$ term as in
Ref.~\cite{Haidenbauer:2017sws}, which generates difference between
interactions of the $J^{P} = 2^{+}$ and $1^{+}$ channels. However, the
term in Eq.~\eqref{eq:Lcontact} is sufficient in this study because we
focus only on the $J^{P} = 2^{+}$ channel.

In the following we construct the $N \Omega$ interaction by using
these effective Lagrangians, together with the empirical information of the two-meson correlation.

\subsection{\boldmath $\eta$ exchange}
\label{sec:2C}

The $\eta$ exchange term, which is denoted by $V_{\rm A}$, can be
straightforwardly formulated according to Fig.~\ref{fig:V_NO}(A).  From
the effective Lagrangians, we can express $V_{\rm A}$ in terms of the
helicity eigenstates as
\begin{align}
  & V_{\rm A} =
  \frac{2 \sqrt{2} ( D - 3 F ) f_{P D D} m_{N} m_{\Omega}}
  {3 f_{\eta} m_{\pi}} 
  \frac{F ( q )^{2}}{q^{2} + m_{\eta}^{2}}
  \notag \\
  & \times 
  \bar{u}_{N} ( \bm{p}^{\prime} , \, \lambda _{N}^{\prime} ) \gamma _{5}
  u_{N} ( \bm{p} , \, \lambda _{N} )
  \bar{u}_{\Omega \, \mu} ( - \bm{p}^{\prime} , \, \lambda _{\Omega}^{\prime} )
   \gamma _{5} u_{\Omega}^{\mu} ( - \bm{p} , \, \lambda _{\Omega} ) ,
\end{align}
where $\bm{p}$ and $\bm{p}^{\prime}$ are the center-of-mass relative
momenta in the initial and final states, respectively, $q \equiv |
\bm{p} - \bm{p}^{\prime} |$ is the momentum transfer, $m_{\eta}$ is
the $\eta$ mass, and $u_{N}$ and $u_{\Omega}^{\mu}$ are the $N$ and
$\Omega$ spinors, respectively (see Appendix~\ref{app:conv}).  The
spinors depends on the helicity $\lambda$ as well as the momentum
$\bm{p}$.  We introduced a form factor $F ( q )$ of a monopole type:
\begin{equation}
  F ( q ) = \frac{\Lambda ^{2}}{\Lambda ^{2} + q^{2}} ,
  \label{eq:FF}
\end{equation}
with a cutoff $\Lambda$.  In the calculation of $V_{\rm A}$ we used
relations
\begin{equation}
  \bar{u}_{N} ( \Slash{p}_{N}^{\prime} - \Slash{p}_{N} ) \gamma _{5} u_{N}
  = 2 m_{N} \bar{u}_{N} \gamma _{5} u_{N} ,
\end{equation}
\begin{equation}
  \bar{u}_{\Omega \, \mu} ( \Slash{p}_{\Omega}^{\prime} - \Slash{p}_{\Omega} )
  \gamma _{5} u_{\Omega}^{\mu}
  = 2 m_{\Omega} \bar{u}_{\Omega \, \mu} \gamma _{5} u_{\Omega}^{\mu} ,
\end{equation}
where $m_{N}$ and $m_{\Omega}$ are the $N$ and $\Omega$ masses,
respectively.

\subsection{Correlated two-meson exchange}
\label{sec:2D}

To formulate the correlated two-meson exchange term $V_{\rm B}$, we
need some consideration.  In this study we start with a general form
of the interaction constructed as a linear combination of the
so-called kinematic covariants $\mathcal{O}_{a}^{(N)}$ and
$\mathcal{O}_{a \, \mu \nu}^{( \Omega )}$~\cite{Kim:1994ce,
  Reuber:1995vc}:
\begin{align}
  V_{\rm B} = \sum _{a} & \mathcal{V}_{a} ( t ) 
  \bar{u}_{N} ( \bm{p}^{\prime} , \, \lambda _{N} ) 
  \mathcal{O}_{a}^{(N)} u_{N} ( \bm{p} , \, \lambda _{N} )
  \notag \\ & \times 
  \bar{u}_{\Omega}^{\mu} ( - \bm{p}^{\prime} , \, \lambda _{\Omega} ) 
  \mathcal{O}_{a \, \mu \nu}^{( \Omega )} 
  u_{\Omega}^{\nu} ( - \bm{p} , \, \lambda _{\Omega} ) ,
\end{align}
where the coefficients $\mathcal{V}_{a}$ are Lorentz-invariant
amplitudes as functions of the Mandelstam variable $t$ and $a$
specifies types of ($\mathcal{O}_{a}^{(N)}$, $\mathcal{O}_{a \, \mu
  \nu}^{( \Omega )}$).  The kinematic covariants are built up from the
Dirac matrices and the momenta in such a way that their bilinear
spinor representations are Lorentz invariant.  Then, because the
correlated two mesons in Fig.~\ref{fig:V_NO} are in the scalar
channel, we have only two independent types of
($\mathcal{O}_{a}^{(N)}$, $\mathcal{O}_{a \, \mu \nu}^{( \Omega )}$),
for which we take
\begin{equation}
  a = \text{S}: ~ 
  \mathcal{O}_{\rm S}^{(N)} = 1, \, 
  \mathcal{O}_{\text{S} \, \mu \nu}^{( \Omega )} = g_{\mu \nu} ,
\end{equation}
\begin{equation}
  a = \text{2M}: ~ 
  \mathcal{O}_{\rm 2M}^{(N)} = 1, \, 
  \mathcal{O}_{\text{2M} \, \mu \nu}^{( \Omega )} 
  = \frac{q_{\mu} q_{\nu}}{m_{\Omega}^{2}} ,
\end{equation}
where $q^{\mu}$ is the four-momentum transfer.  Therefore, making the
three-dimensional reduction and introducing a phenomenological form
factor, we can express $V_{\rm B}$ as
\begin{align}
  V_{\rm B} = & F ( q )^{2}
  \bar{u}_{N} ( \bm{p}^{\prime} , \, \lambda _{N}^{\prime} ) 
  u_{N} ( \bm{p} , \, \lambda _{N} )
  \notag \\
  & \times \left [ \mathcal{V}_{\rm S} ( t )
    \bar{u}_{\Omega \, \mu} ( - \bm{p}^{\prime} , \, \lambda _{\Omega}^{\prime} ) 
  u_{\Omega}^{\mu} ( - \bm{p} , \, \lambda _{\Omega} ) 
  \phantom{\frac{q_{\mu} q_{\nu}}{m_{\Omega}^{2}}} \right .
  \notag \\
  & \phantom{\times [} \left . + \mathcal{V}_{\rm 2M} ( t )
    \frac{q_{\mu} q_{\nu}}{m_{\Omega}^{2}}
  \bar{u}_{\Omega}^{\mu} ( - \bm{p}^{\prime} , \, \lambda _{\Omega}^{\prime} ) 
  u_{\Omega}^{\nu} ( - \bm{p} , \, \lambda _{\Omega} ) \right ] ,
  \label{eq:VB}
\end{align}
where $t = - | \bm{p} - \bm{p}^{\prime} |^{2}$ and $q^{\mu} = ( 0
  , \, \bm{p} - \bm{p}^{\prime})$.  We use the monopole-type form
  factor in Eq.~\eqref{eq:FF} with the same cutoff $\Lambda$.

\begin{figure}[!t]
  \centering
  \PsfigII{0.215}{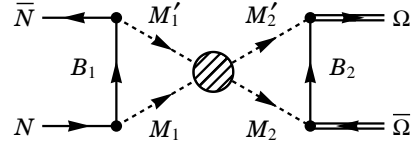}
  \caption{Feynman diagram for the $N \bar{N} \to \Omega \bar{\Omega}$
    reaction.  Particles in the intermediate states are listed in
    Table~\ref{tab:BMM}.  Shaded circle denotes the correlation of two
    mesons.}
  \label{fig:NNbar}
\end{figure}

\begin{table}[!t]
  \caption{Particles in the intermediate states of the diagram in
    Fig.~\ref{fig:NNbar}. }
  \label{tab:BMM}
  \begin{ruledtabular}
    \begin{tabular}{ccc|ccc}
      $B_{1}$ & $M_{1}$ & $M_{1}^{\prime}$
      &
      $B_{2}$ & $M_{2}$ & $M_{2}^{\prime}$
      \\
      \hline
      $N$ & $\pi$ & $\pi$
      &
      $\Omega$ & $\eta$ & $\eta$
      \\
      $N$ & $\eta$ & $\eta$
      &
      $\Xi$ & $K$ & $\bar{K}$
      \\
      $\Delta$ & $\pi$ & $\pi$
      &
      $\Xi ^{\ast}$ & $K$ & $\bar{K}$
      \\
      $\Lambda$ & $K$ & $\bar{K}$
      & & &
      \\
      $\Sigma$ & $K$ & $\bar{K}$
      & & &
      \\
      $\Sigma ^{\ast}$ & $K$ & $\bar{K}$
      & & &
    \end{tabular}
  \end{ruledtabular}
\end{table}

Now our task is to evaluate the coefficients $\mathcal{V}_{\rm S}$ and
$\mathcal{V}_{\rm 2M}$, which govern the interaction strength.  They
are calculated with the dispersion relation for the scattering
amplitude, as done in, e.g., Refs~\cite{Kim:1994ce, Reuber:1995vc}.
In the $N \Omega \to N \Omega$ reaction, $\mathcal{V}_{\text{S},
  \text{2M}} ( t )$ as a function of $t$ is analytic except for some
resonance poles and branch cuts along the real $t$ line: the
unitarity cut running from $4 m_{\pi}^{2}$ to $\infty$ and the
left-hand cuts.  Therefore, neglecting the latter one, which is
irrelevant to the correlated two-meson exchange, we may consider the
dispersion relation in a general form:
\begin{equation}
  \mathcal{V}_{\text{S}, \text{2M}} ( t ) = \frac{1}{\pi}
  \int _{4 m_{\pi}^{2}}^{t_{c}} d t^{\prime}
  \frac{\text{Im} \mathcal{V}_{\text{S}, \text{2M}} ( t^{\prime} )}
       {t^{\prime} - t} ,
  \label{eq:disp}
\end{equation}
where we introduced a cutoff $t_{c}$ instead of infinity.
Equation~\eqref{eq:disp} means that, to calculate the $N \Omega$
interaction with the correlated two-meson exchange taking place in the
region $t < 0$, we may consider the same amplitude but in $t > 4
m_{\pi}^{2}$, which can be achieved in the $N \bar{N} \to \Omega
\bar{\Omega}$ reaction as shown in Fig.~\ref{fig:NNbar}.

In this study we formulate the $N \bar{N} \to \Omega \bar{\Omega}$
scattering amplitude by considering the intermediate states listed in
Table~\ref{tab:BMM}.  The scattering amplitude of the two pseudoscalar
mesons in the scalar channel, denoted by the shaded circle in
Fig.~\ref{fig:NNbar}, is calculated in the so-called chiral unitary
approach~\cite{Oller:1997ti, Oller:1997ng, Oller:1998hw,
  Oller:1998zr}.  The details of the formulation and calculation of
the $N \bar{N} \to \Omega \bar{\Omega}$ scattering amplitude are given
in Appendix~\ref{app:corr}.

\begin{figure}[!t]
  \centering
  \Psfig{8.6cm}{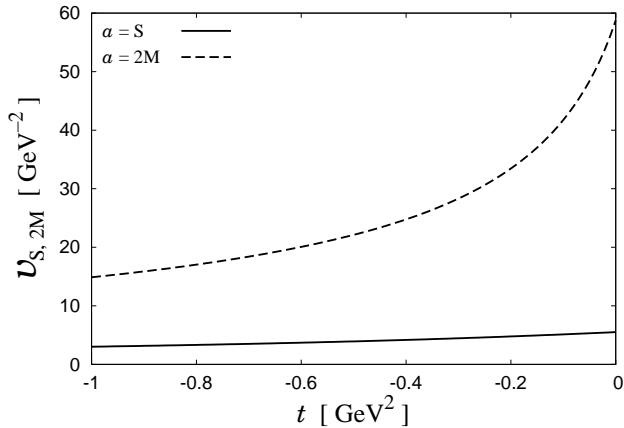}
  \caption{Lorentz invariant amplitudes $\mathcal{V}_{\rm S}$ and
    $\mathcal{V}_{\rm 2M}$ calculated with the dispersion relation as
    functions of the Mandelstam variable $t$.}
  \label{fig:VS2M}
\end{figure}

We fix the cutoff $t_{c} = ( 1.2 \gev )^{2}$, which is the upper
boundary of the fit range of our $\pi \pi$-$K \bar{K}$-$\eta \eta$
scattering amplitude in the chiral unitary approach to the
experimental $\pi \pi ( J = 0 , \, I = 0 )$ phase shift
(Appendix~\ref{app:meson}).  The resulting $\mathcal{V}_{\text{S},
  \text{2M}}$ in the region $t < 0$ are shown in Fig.~\ref{fig:VS2M}.

\subsection{Contact term}
\label{sec:2E}

The contact term is straightforwardly constructed as
\begin{align}
  V_{\rm C} = & - c F ( q )^{2} 
  \bar{u}_{N} ( \bm{p}^{\prime} , \, \lambda _{N}^{\prime} ) 
  u_{N} ( \bm{p} , \, \lambda _{N} )
  \notag \\ 
  & \times \bar{u}_{\Omega \, \mu} ( - \bm{p}^{\prime} , \,
  \lambda _{\Omega}^{\prime} ) 
  u_{\Omega}^{\mu} ( - \bm{p} , \, \lambda _{\Omega} ) .
\end{align}
We here introduced the form factor $F ( q )$ as in the $\eta$ and
correlated two-meson exchanges.  The unknown coupling constant $c$ is
to be determined by the lattice QCD data as described in
Sec.~\ref{sec:3}.

\subsection{Projection to partial waves and coupling to inelastic channels}
\label{sec:2F}

The interaction terms above are constructed in terms of the helicity
eigenstates in momentum space as
\begin{equation}
  V = V ( \bm{p}^{\prime} , \, \lambda _{3} , \, \lambda _{4}
  , \, \bm{p} , \, \lambda _{1} , \, \lambda _{2} ) ,
  \label{eq:Vgeneral}
\end{equation}
where $\bm{p}$ and $\bm{p}^{\prime}$ are the 
relative momenta in the center-of-mass frame in the initial and
final states, respectively, and $\lambda _{a}$ is the helicity for the
$a$th baryon in the baryon--baryon scattering $B_{1} B_{2} \to B_{3}
B_{4}$.  Now these interaction terms are projected to partial waves
according to the method in Appendix~\ref{app:proj}. As a result, the
expression of the interaction reduces to
\begin{equation}
  V = V_{\alpha} ( p^{\prime} , \, p ),
\end{equation}
with $p^{(\prime)}=|\bm{p}^{(\prime)}|$ and $\alpha = ( J , \, P , \,
L^{\prime} , \, S^{\prime} , \, L , \, S )$, where $J$, $P$,
  $L^{( \prime )}$, and $S^{( \prime )}$ are total angular momentum,
  parity, spin, and orbital angular momentum of the two-body system in
  the initial (final) state, respectively.

\begin{table}
  \caption{Channels for the $N \Omega$ coupled-channels scattering. }
  \label{tab:chanCC}
  \begin{ruledtabular}
    \begin{tabular}{cc}
      1 & $N \Omega$ (${}^{5} S_{2}$) 
      \\
      2 & $\Lambda \Xi$ (${}^{3} D_{2}$) 
      \\
      3 & $\Lambda \Xi$ (${}^{1} D_{2}$) 
      \\
      4 & $\Sigma \Xi$ (${}^{3} D_{2}$) 
      \\
      5 & $\Sigma \Xi$ (${}^{1} D_{2}$) 
      \\
      6 & $\Lambda \Xi ^{\ast}$ (${}^{5} S_{2}$) 
    \end{tabular} 
  \end{ruledtabular}
\end{table}

In this study we focus on the $N \Omega$ interaction in its $S$ wave
with $J^{P} = 2^{+}$ (${}^{5}S_{2}$ channel) where the attractive
interaction is reported by the HAL QCD collaboration.  Let us discuss
contributions from the inelastic channels $\Lambda \Xi$, $\Sigma \Xi$,
and $\Lambda \Xi ^{\ast}$.  Here we consider these channels with
minimal orbital angular momenta, namely, $D$, $D$, and $S$ waves for
the $\Lambda \Xi$, $\Sigma \Xi$, and $\Lambda \Xi ^{\ast}$ channels,
respectively.  We note that for $J^{P} = 2^{+}$ there are two $D$-wave
states with different spins (${}^{3} D_{2}$ and ${}^{1} D_{2}$) in
each of the $\Lambda \Xi$ and $\Sigma \Xi$ channels.  In summary, we
take into account the channels listed in Table~\ref{tab:chanCC}.

The evaluation of the inelastic contributions proceeds as follows.  We
first calculate a coupled-channel partial-wave projected interaction
of the process $j ( p ) \to 1 ( p^{\prime} )$, $V_{1 j} ( p^{\prime} ,
\, p )$ where the channel index $j$ runs from $1$ to $6$ as listed in
Table~\ref{tab:chanCC}.  Owing to the time reversal invariance of the
strong interaction, we have a relation $ V_{1 j} ( p^{\prime} , \, p )
= V_{j 1} ( p , \, p^{\prime} ) $.  The transition to the inelastic
channels is driven by the $K$ exchange as described in
Sec.~\ref{sec:2G}.  We then evaluate the box diagram in
Section~\ref{sec:2H} to obtain the effective single-channel $N \Omega$
interaction in channel 1.\footnote{The $N \Omega({}^{5}D_{2})$ and $N
  \Omega({}^{3}D_{2})$ channels couple to this sector through the
  tensor force in the $\eta$ exchange. We will estimate this effect in
  Sec.~\ref{sec:4A} and show that their contributions are small.}

\subsection{\boldmath $K$ exchange for transition interaction}
\label{sec:2G}

The transition to inelastic channels with the $K$ exchange as shown in
Fig.~\ref{fig:V_NO}(box) can be formulated in a similar manner to the
case of the $\eta$ exchange.  Here, for simplicity we make an
approximation that the time component of the momentum transfer is
zero, $q^{0} \approx 0$, and hence $q^{\mu} \approx ( 0 , \, \bm{p} -
\bm{p}^{\prime} )$.  This can be guaranteed by the mass degeneracy in
the $\SUN{6}$ spin-flavor symmetry.  Then, the transition terms are
constructed as
\begin{align}
  & V_{N \Omega \to \Lambda \Xi} =  
  - \frac{( D + 3 F ) f_{P B D} ( m_{N} + m_{\Lambda} )}
  {2 \sqrt{3} f_{K} m_{\pi}} 
  \frac{F ( q )^{2}}{q^{2} + m_{K}^{2}}
  \notag \\
  & \times 
  \bar{u}_{\Lambda} ( \bm{p}^{\prime} , \, \lambda _{\Lambda}^{\prime} ) \gamma _{5}
  u_{N} ( \bm{p} , \, \lambda _{N} )
  \bar{u}_{\Xi} ( - \bm{p}^{\prime} , \, \lambda _{\Xi}^{\prime} )
  q_{\mu} u_{\Omega}^{\mu} ( - \bm{p} , \, \lambda _{\Omega} ) ,
\end{align}
\begin{align}
  & V_{N \Omega \to \Sigma \Xi} =  
  - \frac{\sqrt{3} ( D - F ) f_{P B D} ( m_{N} + m_{\Sigma} )}
  {2 f_{K} m_{\pi}} 
  \frac{F ( q )^{2}}{q^{2} + m_{K}^{2}}
  \notag \\
  & \times 
  \bar{u}_{\Sigma} ( \bm{p}^{\prime} , \, \lambda _{\Sigma}^{\prime} ) \gamma _{5}
  u_{N} ( \bm{p} , \, \lambda _{N} )
  \bar{u}_{\Xi} ( - \bm{p}^{\prime} , \, \lambda _{\Xi}^{\prime} )
  q_{\mu} u_{\Omega}^{\mu} ( - \bm{p} , \, \lambda _{\Omega} ) ,
\end{align}
\begin{align}
  & V_{N \Omega \to \Lambda \Xi ^{\ast}} 
  = - \frac{( D + 3 F ) f_{P D D} ( m_{N} + m_{\Lambda} ) 
    ( m_{\Xi ^{\ast}} + m_{\Omega} )}
  {6 f_{K} m_{\pi}} 
  \notag \\ & \times 
  \frac{F ( q )^{2}}{q^{2} + m_{K}^{2}}
  \bar{u}_{\Lambda} ( \bm{p}^{\prime} , \, \lambda _{\Lambda}^{\prime} ) \gamma _{5}
  u_{N} ( \bm{p} , \, \lambda _{N} )
  \notag \\
  & \times 
  \bar{u}_{\Xi ^{\ast} \, \mu} ( - \bm{p}^{\prime} , \, 
  \lambda _{\Xi ^{\ast}}^{\prime} )
  \gamma _{5} u_{\Omega}^{\mu} ( - \bm{p} , \, \lambda _{\Omega} ) ,
\end{align}
with $q \equiv | \bm{q} | = | \bm{p} - \bm{p}^{\prime} |$.  We adopt
the same form factor $F(q)$ with the other diagrams. Performing the
partial wave projection in Appendix~\ref{app:proj}, we obtain
$V_{1 j}(p^{\prime}, \, p)$.

\subsection{Inelastic contributions in box diagrams}
\label{sec:2H}

As we explained in Sec.~\ref{sec:2A}, we consider the transition of
the $N \Omega$ to the inelastic channels by the $K$ exchange but
neglect the transition between the inelastic channels such as $\Lambda
\Xi \to \Lambda \Xi$.  In this case, the inelastic channels contribute
to the $N \Omega ( {}^{5} S_{2} )$ interaction only through the box
diagrams in Fig.~\ref{fig:V_NO}(box).  We can express this by using
the partial-wave projected interaction in the previous section:
\begin{equation}
  V_{\text{box}(j)} ( E ; \, p^{\prime} , p )
  = \int _{0}^{\infty} \frac{d p^{\prime \prime}}{2 \pi ^{2}} p^{\prime \prime \, 2}
  \frac{V_{1 j} ( p^{\prime} , \, p^{\prime \prime} )
    V_{j 1} ( p^{\prime \prime} , \, p )}
       {E - \mathcal{E}_{j} ( p^{\prime \prime} ) + i 0} ,
       \label{eq:Vbox}
\end{equation}
with $j = 2$--$6$.  The on-shell energy $\mathcal{E}_{j}$ is
\begin{equation}
  \mathcal{E}_{j} ( p )
  \equiv \sqrt{p^{2} + m_{j}^{2}} + \sqrt{p^{2} + m_{j}^{\prime \, 2}} 
\end{equation}
with $m_{j}$ and $m_{j}^{\prime}$ being masses of particles in channel
$j$: $( m_{2}, \, m_{2}^{\prime}) = ( m_{\Lambda} , \, m_{\Xi})$, $(
m_{6}, \, m_{6}^{\prime}) = ( m_{\Lambda} , \, m_{\Xi ^{\ast}})$, etc.
We note that the interaction $V_{\text{box}(j)} ( E ; \, p^{\prime} ,
p )$ depends on the center-of-mass energy $E$.  For a real energy $E >
m_{i} + m_{i}^{\prime}$, the interaction $V_{\text{box}(j)} ( E ; \,
p^{\prime} , p )$ becomes complex according to the infinitesimal
quantity $+ i 0$ in the denominator.  The imaginary part of the box
interaction $V_{\text{box}(j)}$ represents the absorption of the $N
\Omega$ system into the channel $j$.

\section{\boldmath Scattering amplitude and parameter fixing}
\label{sec:3}

The $N \Omega ( {}^{5} S_{2} ) $ interaction we have formulated is
composed of
\begin{align}
  V ( E ; \, p^{\prime} , \, p )
  = & V_{\rm A} ( p^{\prime} , \, p )
  + V_{\rm B} ( p^{\prime} , \, p )
  + V_{\rm C} ( p^{\prime} , \, p )
  \notag \\
  & 
  + \sum _{j = 2}^{6} V_{\text{box}(j)} ( E ; \, p^{\prime} , \, p ) .
  \label{eq:Vfull}
\end{align}
Here $V_{\rm A,B,C}$ denote the contributions from the $\eta$
exchange, correlated two-meson exchange, and contact terms projected
to the ${}^{5} S_{2}$ channel, respectively.  The box contribution
$V_{\text{box} (j)}$ was defined in Eq.~\eqref{eq:Vbox}.

One of the most important quantities calculated with this interaction
is the $T$-matrix of the $N \Omega ( {}^{5} S_{2} )$ scattering.  In
the present formulation, the $T$-matrix of the $N \Omega ( {}^{5}
S_{2} )$ scattering is a solution of the Lippmann--Schwinger equation
in a single channel as follows:
\begin{align}
  & T ( E ; \, p^{\prime} , \, p )
  = V ( E ; \, p^{\prime} , \, p )
  \notag \\
  & \quad
  + \int _{0}^{\infty}
  \frac{d p^{\prime \prime}}{2 \pi ^{2}} p^{\prime \prime \, 2}
  \frac{V ( E ; \, p^{\prime} , \, p^{\prime \prime} )
    T ( E ; \, p^{\prime \prime} , \, p )}
       {E - \mathcal{E}_{N \Omega} ( p^{\prime \prime} ) + i 0} ,
       \label{eq:LSeq_full}
\end{align}
where $\mathcal{E}_{N \Omega}$ is the on-shell energy for the $N
\Omega$ system
\begin{equation}
  \mathcal{E}_{N \Omega} ( p )
  \equiv \sqrt{p^{2} + m_{N}^{2}} + \sqrt{p^{2} + m_{\Omega}^{2}} .
  \label{eq:Ephys}
\end{equation}
The energy $E$ in Eq.~\eqref{eq:LSeq_full} can be analytically
continued to the complex plane.  When we calculate the on-shell
$T$-matrix, $T_{\rm on}$, for the $N \Omega ( {}^{5} S_{2} )$
scattering above the threshold, the infinitesimal quantity $+ i 0$ in
the denominator specifies the boundary condition and gives the
imaginary part of the $T$-matrix, which results in
\begin{equation}
  T_{\text{on}} ( k )
  = \frac{K_{\text{on}} ( k )}
  {1 + K_{\text{on}} ( k ) \times i \rho ( k ) / 2} ,
  \label{eq:Ton}
\end{equation}
with the relative momentum $k$, on-shell $K$-matrix $K_{\text{on}}$,
and phase space $\rho ( k )$.  The $K$-matrix is calculated with the
integral equation
\begin{align}
  & K ( E ; \, p^{\prime} , \, p )
  = V ( E ; \, p^{\prime} , \, p )
  \notag \\
  & \quad
  + \mathcal{P} \int _{0}^{\infty}
  \frac{d p^{\prime \prime}}{2 \pi ^{2}} p^{\prime \prime \, 2}
  \frac{V ( E ; \, p^{\prime} , \, p^{\prime \prime} )
    K ( E ; \, p^{\prime \prime} , \, p )}
       {E - \mathcal{E}_{N \Omega} ( p^{\prime \prime} )} ,
  \label{eq:Kmat}
\end{align}
where $\mathcal{P}$ stands for the principal value of the integral, and the on-shell part is
\begin{equation}
  K_{\text{on}} ( k ) \equiv
  K ( \mathcal{E}_{N \Omega} ( k ) ; \, k , \, k ) .
  \label{eq:Kon}
\end{equation}
The phase space $\rho ( k )$ is defined as
\begin{equation}
  \rho ( k ) \equiv \frac{k}{\pi}
  \frac{\sqrt{\left ( k^{2} + m_{N}^{2} \right )
      \left ( k^{2} + m_{\Omega}^{2} \right )}}{\mathcal{E}_{N \Omega} ( k )} .
  \label{eq:rho_PS}
\end{equation}

From the on-shell $T$-matrix, we can extract the threshold parameters
for the $N \Omega$ scattering in nonrelativistic quantum mechanics
such as the scattering length $a$ and effective range $r_{\rm eff}$.
In the present notation, the $N \Omega ( {}^{5} S_{2} )$ scattering
amplitude in nonrelativistic quantum mechanics, $f_{S}$, is expressed
as
\begin{equation}
  f_{S} ( k ) = - \frac{1}{2 \pi}
  \frac{\sqrt{\left ( k^{2} + m_{N}^{2} \right )
    \left ( k^{2} + m_{\Omega}^{2} \right )}}{\mathcal{E}_{N \Omega} ( k )}
  T_{\text{on}} ( k ) .
  \label{eq:fS}
\end{equation}
We can expand the inverse of the scattering amplitude, $f_{S} ( k
)^{-1}$, with respect to the relative momentum $k$ as
\begin{equation}
  f_{S} ( k )^{-1} = - \frac{1}{a} - i k + \frac{1}{2} r_{\rm eff} k^{2} 
  + \mathcal{O} ( k^{4} ) ,
  \label{eq:erange_exp}
\end{equation}
where the scattering length $a$ and effective range $r_{\rm eff}$
enter as the coefficients of the zeroth and second order terms,
respectively.  Therefore, we can calculate them from the behavior of
the scattering amplitude at the threshold:
\begin{equation}
  a = - f_{S} ( k = 0 ) , 
  \label{eq:scatt_length_full}
\end{equation}
\begin{equation}
  r_{\rm eff} = \left [ \frac{d^{2} f_{S}^{-1}}{d k^{2}} \right ]_{k = 0} .
  \label{eq:reff_full}
\end{equation}
We note that the scattering length and effective range are in general
complex if the interaction $V$ has imaginary part as the absorption
into open channels.

Next, we would like to fix the model parameters in our potential:
cutoff $\Lambda$ and coupling constant for the contact term $c$.
Among the two parameters, the cutoff $\Lambda$ can be fixed to be a
typical hadron scale.  In the present study we take the value $\Lambda
= 1.0 \gev$.  The coupling constant $c$, on the other hand, we need
information on the $N \Omega ( {}^{5} S_{2} )$ interaction.  We employ
the recent HAL QCD result on the scattering length from lattice QCD
simulations with the nearly physical quark masses~\cite{Doi:2017zov},
in which the $N \Omega ( {}^{5}S_{2} )$ system is found to be bound
with very small binding energy and the system is almost in the unitary
limit.  In Ref.~\cite{Doi:2017zov}, they reported that the scattering
length is $7.4 \pm 1.6 \fm$ at the time range $t = 11$ of the lattice
simulations.\footnote{We would like to thank T.~Iritani and HAL QCD
  collaboration for providing us with the numerical value of the
  scattering length~\cite{Iritani:2018b}.  The HAL QCD collaboration
  provides a real-valued scattering length because of $N \Omega (
  {}^{5} S_{2} )$ single-channel analysis.  We also note that the
  scattering length in our notation [see Eq.~\eqref{eq:erange_exp}]
  has opposite sign to that in HAL QCD.}  We reproduce this value by
using our model but with hadron masses adjusted to the lattice
simulations.

For the hadron masses in the lattice QCD simulations, we adopt
$m_{N}^{\rm lat} = 964 \mev$ and $m_{\Omega}^{\rm lat} = 1712 \mev$
taken from~\cite{Gongyo:2017fjb}, $m_{\Lambda}^{\rm lat} = 1123 \mev$,
$m_{\Sigma}^{\rm lat} = 1204 \mev$, and $m_{\Xi}^{\rm lat} = 1332
\mev$ from~\cite{Sasaki:2017ysy}, and $m_{\Xi ^{\ast}}^{\rm lat} =
1580 \mev$ from Fig.~1 of~\cite{Ishikawa:2015rho}.  They are used to
calculate the on-shell energies $\mathcal{E}_{j}$, which enter in the
denominators of the $T$-matrix and box interactions, while we assume
that the interaction~\eqref{eq:Vfull} remains unchanged.  In addition,
to simulate the contributions from the $\Lambda \Xi$ and $\Sigma \Xi$
channels in finite volume in our framework, we take the real part of
the box interaction $V_{\text{box} (j)}$.  Then, the scattering length
is calculated as in Eq.~\eqref{eq:scatt_length_full} together with
Eqs.~\eqref{eq:Ton}--\eqref{eq:fS} but with masses being $m_{N}^{\rm
  lat}$ and $m_{\Omega}^{\rm lat}$ instead of $m_{N}$ and
$m_{\Omega}$, respectively.

In this condition, we obtain the scattering length $a = 7.4 \fm$ with
the coupling constant $c = - 22.1 \gev ^{-2}$.  In the followings, we
adopt this value of the coupling constant $c = - 22.1 \gev ^{-2}$
together with the cutoff $\Lambda = 1.0 \gev$.  We note that, if we
keep the hadron masses in the lattice simulations but turn on the
imaginary part of the box interaction, we obtain the scattering length
$a = 4.1 - 3.1 i \fm$.

\section{\boldmath Properties of the $N \Omega$ interaction}
\label{sec:4}

Now that we have fixed parameters in our model, we discuss the
properties of the $N \Omega ({}^{5} S_{2})$ interaction.  In the
following, we use the physical hadron masses in the isospin-symmetric
limit summarized in Appendix~\ref{app:masses}.

\subsection{Elastic contributions}
\label{sec:4A}

\begin{figure}[!t]
  \centering
  \Psfig{8.6cm}{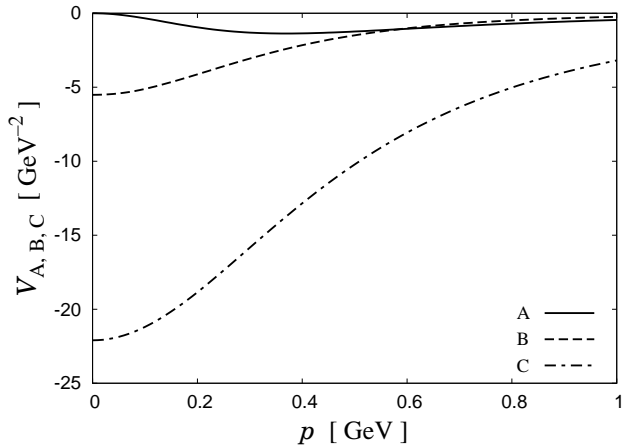}
  \caption{The elastic contributions to the $N \Omega ( ^{5} S_{2} )$
    interaction $V_{\rm A,B,C} ( p , \, p )$ as functions of momentum
    $p$.}
  \label{fig:Vpp}
\end{figure}

First we investigate properties of the $N \Omega ({}^{5} S_{2})$
interaction in the elastic $N \Omega$ channel, i.e., the $V_{\rm
  A,B,C} ( p^{\prime} , \, p )$ terms in Eq.~\eqref{eq:Vfull}.  We
note that the terms $V_{\rm A,B,C}$ has no dependence on the energy
$E$.  The contributions $V_{\rm A,B,C} (p^{\prime} =p, \, p)$ are
plotted in Fig.~\ref{fig:Vpp} as functions of the momentum $p$.  As
one can see from the figure, the contact term $V_{\rm C}$ is strongly
attractive and gives a dominant contribution.  Other two terms,
$V_{\rm A}$ and $V_{\rm B}$, give moderate attraction on top of the
contact term.  This finding of the weak $\eta$ and ``$\sigma$''
exchanges is consistent with the calculation based on a quark model in
Ref.~\cite{Huang:2015yza}.  The $\eta$ exchange interaction $V_{\rm
  A}$ is weak because the $\eta N N$ coupling constant is small.  The
correlated two-meson exchange interaction $V_{\rm B}$ is also weak.
This is in contrast to the $N N$ case, in which the broad ``$\sigma$
meson'' plays an important role to generate $N N$ attraction in the
intermediate range region.  The weakness of the correlated two-meson
exchange in the $N\Omega$ system is because the broad ``$\sigma$
meson'' cannot couple to the $\Omega$ via the $\pi \pi$ state [see
  Eq.~\eqref{eq:pipiOmegaOmega}].  Another resonance in the
scalar-isoscalar channel, $f_{0} (980)$, has been considered to be a
$K \bar{K}$ molecular state~\cite{Baru:2003qq, Sekihara:2014kya,
  Sekihara:2014qxa} and hence it can couple both to $N$ and $\Omega$.
However, this contribution turns out to be also small, presumably
because the heavier meson exchange acts only at the short range (high
momentum) region.

\begin{table}[t]
 \caption{Volume integral~\eqref{eq:volume} of the $N \Omega$
   interaction from each contribution in units of $\gev ^{-1}$.  The
   energy is fixed as $E = 2550 \mev$, $m_{N} + m_{\Omega} = 2611.4
   \mev$, and $2650 \mev$.  Contributions from inelastic channels are
   evaluated as the box terms.}
  \label{tab:volume}
  \begin{ruledtabular}
    \begin{tabular}{lrrr}
      Contribution
      & \multicolumn{1}{c}{$2550 \mev$} &
      \multicolumn{1}{c}{$m_{N} + m_{\Omega}$} &
      \multicolumn{1}{c}{$2650 \mev$} \\
      \hline
      A & $-1.11 \phantom{- 0.00 ii}$ &
      $-1.11 \phantom{- 0.00 ii}$ &
      $-1.11 \phantom{- 0.00 ii}$
      \\
      B & $-2.22 \phantom{- 0.00 ii}$ &
      $-2.22 \phantom{- 0.00 ii}$ &
      $-2.22 \phantom{- 0.00 ii}$ 
      \\
      C & $-13.21 \phantom{- 0.00 ii}$ &
      $-13.21 \phantom{- 0.00 ii}$ &
      $-13.21 \phantom{- 0.00 ii}$
      \\
      $N \Omega ( {}^{5} D_{2} ) $ &
      $-0.08 \phantom{- 0.00 ii}$ &
      $-0.10 \phantom{- 0.00 ii}$ &
      $-0.12 - 0.01 i$
      \\
      $N \Omega ( {}^{3} D_{2} ) $ & 
      $-0.03 \phantom{- 0.00 ii}$ & 
      $-0.03 \phantom{- 0.00 ii}$ & 
      $-0.04 - 0.00 i$
      \\
      $\Lambda \Xi ({}^{3} D_{2})$ & $-1.41 - 0.55 i$ & $-1.32 - 0.94 i$ 
      & $-1.19 - 1.13 i$
      \\
      $\Lambda \Xi ({}^{1} D_{2})$ & $-0.92 - 0.37 i$ & $-0.86 - 0.62 i$ 
      & $-0.78 - 0.75 i$ 
      \\
      $\Sigma \Xi ({}^{3} D_{2})$ & $-0.24 - 0.02 i$ & $-0.28 - 0.09 i$ 
      & $-0.28 - 0.15 i$ 
      \\
      $\Sigma \Xi ({}^{1} D_{2})$ & $-0.16 - 0.01 i$ & $-0.18 - 0.06 i$ 
      & $-0.18 - 0.10 i$
      \\
      $\Lambda \Xi ^{\ast} ({}^{5} S_{2})$ &
      $-0.53\phantom{ - 0.00 ii}$ &
      $-0.67 \phantom{- 0.00 ii}$ 
      & $-0.97 - 0.05 i$
      \\
      Total &
      $-19.89 - 0.95 i$ &
      $-19.97 - 1.72 i$ &
      $-20.08 - 2.18 i $
    \end{tabular}
  \end{ruledtabular}
\end{table}

In order to estimate the strength of the attraction, we calculate the
volume integral of the interaction in the momentum space:
\begin{equation}
  \int _{0}^{\infty} d p \, V ( p, \, p ) .
  \label{eq:volume}
\end{equation}
The numerical results of the volume integrals of $V_{\rm A,B,C}$ are
listed in the second, third, and fourth rows in
Table~\ref{tab:volume}, respectively.  We can see that the contact
term (C) is about ten times more attractive than the $\eta$ or
correlated two-meson exchange. In other words, the lattice QCD
scattering length~\cite{Doi:2017zov} requires such attractive
component represented by the contact term, in addition to the
conventional meson exchanges at long distance.

\begin{figure}[!t]
  \centering
  \Psfig{8.6cm}{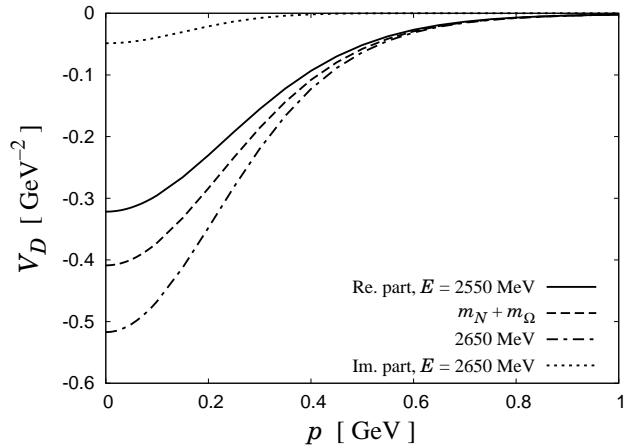}
  \caption{The $D$-wave contribution to the $N \Omega ( {}^{5} S_{2}
    )$ interaction $V_{D} ( E ; \, p , \, p )$ as a function of
    momentum $p$.  The energy in the effective interaction is fixed as
    $E = m_{N} + m_{\Omega}$, $2550 \mev$, and $2650 \mev$.  Because
    the energy is above the $N \Omega$ threshold, the contribution at
    $E = 2650 \mev$ has an imaginary part.}
  \label{fig:VD}
\end{figure}

In addition to the $S$ wave, we examine the $D$-wave $N \Omega$
contribution as well because the $\eta$ exchange term $V_{\rm A}$ can
mix the $S$- and $D$-wave states owing to the tensor-force coupling,
which is essential in the $N N$ system through the $\pi$ exchange.  We
note that there are two $D$-wave states with different spins in the
$J^{P} = 2^{+}$ state, $N \Omega ( {}^{5} D_{2} )$ and $N \Omega (
{}^{3} D_{2} )$, to which we assign the channels $j = 7$ and $8$,
respectively.  We calculate the $D$-wave contribution to the $S$ wave
in the $N \Omega$ system through the box diagrams with intermediate
state being the $N \Omega ( {}^{5} D_{2} )$ and $N \Omega ( {}^{3}
D_{2} )$ channels:
\begin{align}
  & V_{D} ( E ; \, p^{\prime} , \, p )
  = \sum _{j=7}^{8} \int _{0}^{\infty}
  \frac{d p^{\prime \prime}}{2 \pi ^{2}} p^{\prime \prime \, 2}
  \frac{V_{1 j} ( p^{\prime} , \, p^{\prime \prime} )
    V_{j 1} ( p^{\prime \prime} , \, p )}
       {E - \mathcal{E}_{N \Omega} ( p^{\prime \prime} ) + i 0} .
       \label{eq:Vbox_D}
\end{align}
For the interaction $V_{j 1}$ ($j = 7$ and $8$) in the numerator of
the integrand, we consider only the $N \Omega$ channel, $V_{\text{A} +
  \text{B} + \text{C}}$, projected to the $S$ and $D$ waves in the
initial and final states, respectively.  We note that the effective
interaction $V_{D}$ depends on the energy owing to the reduction of
the $D$-wave channels.

The $D$-wave contribution to the $S$-wave interaction $V_{D} ( E; \,
p, \, p )$ is plotted in Fig.~\ref{fig:VD} as a function of the
momentum $p$.  We fix the energy in the effective interaction as $E =
m_{N} + m_{\Omega}=2611.4\mev$, $2550 \mev$, and $2650 \mev$.  Note
that the box term provides an imaginary part of the interaction above
the threshold ($E = 2650 \mev$).  Comparing the result in
Fig.~\ref{fig:VD} with those in Fig.~\ref{fig:Vpp}, we find that the
$D$-wave contribution ($\sim-0.4$ GeV$^{-2}$ at $p=0$ GeV) is very
tiny with respect to the $S$-wave contact term ($\sim-22$ GeV$^{-2}$
at $p=0$ GeV) and hence the $D$ wave contribution in the $S$-wave
interaction is negligible.  We can understand this behavior by the
weak $\eta N N$ coupling compared to the $\pi N N$ coupling.  We also
find that the energy dependence of the $D$-wave contribution in the
$S$-wave interaction is not significant.

To quantify the smallness of the $D$-wave contribution, we calculate
the volume integral~\eqref{eq:volume} of $V_{D}$ as listed in the
fifth and sixth rows in Table~\ref{tab:volume}.  The volume integral
from the $D$-wave contribution is only $\sim 1 \%$ of the contact-term
contribution.

Based on these results, in the following discussions we neglect
the $N \Omega ( {}^{5} D_{2} )$ and $N \Omega ( {}^{3} D_{2} )$
channels.

\subsection{Inelastic contributions}
\label{sec:4B}

\begin{figure}[!t]
  \centering
  \Psfig{8.6cm}{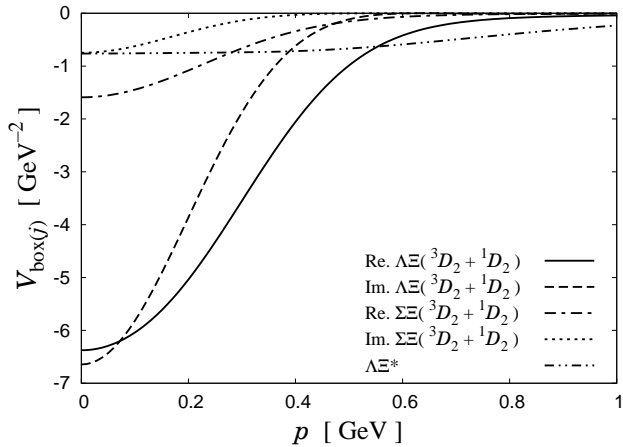}
  \caption{The inelastic contributions to the $N \Omega ( ^{5} S_{2}
    )$ interaction $V_{\text{box} (j)} ( E ; \, p , \, p )$ as
    functions of momentum $p$.  The energy in the effective
    interaction is fixed as $E = m_{N} + m_{\Omega}$.  In the figure
    $\Lambda \Xi$ and $\Sigma \Xi$ indicates the sum of ${}^{3} D_{2}$
    and ${}^{1} D_{2}$ contributions, respectively.}
  \label{fig:Vbox}
\end{figure}

Next we investigate the effects of the inelastic channels to the $N
\Omega ({}^{5} S_{2})$ interaction.

The contributions $V_{\text{box}(j)} ( E ; \, p^{\prime} = p, \, p )$
are plotted in Fig.~\ref{fig:Vbox} as functions of the momentum $p$.
We here show the sum of the ${}^{3} D_{2}$ and ${}^{1} D_{2}$
contributions in the $\Lambda \Xi$ and $\Sigma \Xi$ channels, for
simplicity.  The energy is fixed at threshold $E = m_{N} +
m_{\Omega}$, thus the interaction which involves open $\Lambda \Xi$ or
$\Sigma \Xi$ channel in the intermediate state has an imaginary part.
From the real part of the interaction, we observe that the $\Lambda
\Xi$, $\Sigma \Xi$, and $\Lambda \Xi ^{\ast}$ channels assist the
attraction of the $N \Omega ( {}^{5} S_{2} )$ interaction.  Among
them, the $\Lambda \Xi$ channel gives the strongest attraction, which
is comparable to the correlated two-meson exchange $V_{\rm B}$ (see
Fig.~\ref{fig:Vpp}).  Even with the smaller energy denominator, the
interaction of the intermediate $\Sigma \Xi$ channel is suppressed
compared to the $\Lambda \Xi$ case by the smaller $K N \Sigma$
coupling: $( D + 3 F ) / 2 \sqrt{3} \approx 0.63$ for the $K N
\Lambda$ coupling, and $\sqrt{3} ( D - F ) / 2 \approx 0.29$ for the
$K N \Sigma$ coupling.  The intermediate $\Lambda \Xi ^{\ast}$ channel
becomes significant only at higher momentum $p \gtrsim 0.6 \gev$.  As
for the imaginary part of the interaction, the intermediate $\Lambda
\Xi$ term gives larger contribution than the $\Sigma \Xi$ one, which
indicates the $N \Omega ( {}^{5} S_{2} )$ system mainly decays to the
$\Lambda \Xi$ channel.

We calculate the volume integral~\eqref{eq:volume} from the inelastic
contributions, and the results are listed from the seventh to eleventh
rows in Table~\ref{tab:volume}.  We can see that the $\Lambda \Xi$
channel gives the strongest attraction and absorption among the
inelastic channels.  The sum of the real parts of the volume integrals
from the $\Lambda \Xi({}^{3} D_{2})$ and $\Lambda \Xi({}^{1} D_{2})$
contributions is similar magnitude to the volume integral from the
correlated two-meson exchange ($-2.22 \gev ^{-1}$).  The imaginary
part grows as the energy $E$ increases because a larger phase space
can be utilized for a higher energy $E$.  We can also understand from
Table~\ref{tab:volume} that the energy dependence of the box
interaction is not significant.  Indeed, when we vary the energy from
$E = m_{N} + m_{\Omega}$ to $2550 \mev$ or $2650 \mev$, the shift of
the volume integral in each contribution is only $\lesssim 1 \%$ of
the total amount of the volume integral listed in the last row of
Table~\ref{tab:volume}.

\subsection{\boldmath On-shell $N \Omega ( {}^{5} S_{2} )$
  scattering amplitude}
\label{sec:4C}

We then calculate the on-shell $N \Omega ( {}^{5} S_{2} )$
scattering amplitude above the $N \Omega$ threshold and extract the
scattering length and effective range.

\begin{figure}[!t]
  \centering
  \Psfig{8.6cm}{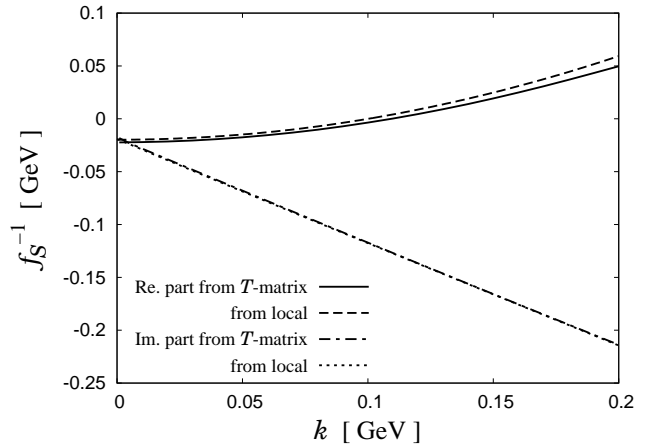}
  \caption{Inverse of the scattering amplitude $f_{S}^{-1}$.  In
    addition to $f_{S}^{-1}$ calculated from the $T$-matrix, we also
    plot $f_{S}^{-1}$ from the equivalent local $N \Omega$ potential
    (see Sec.~\ref{sec:5}).  We note that the two lines of the
    imaginary parts almost overlap each other.}
  \label{fig:fSinv}
\end{figure}

The $N \Omega$ scattering amplitude $f_{S} ( k )$ as a function of the
relative momentum $k$ is obtained by the formula~\eqref{eq:fS}.
Because the inverse of the scattering amplitude $f_{S} ( k )^{-1}$ is
useful to extract the scattering length and effective range, we show
the result of the inverse of the scattering amplitude $f_{S} ( k
)^{-1}$ in Fig.~\ref{fig:fSinv} (solid and dash-dotted lines).
Because the $N \Omega$ interaction is complex reflecting the
absorption into open channels, $\Lambda \Xi$ and $\Sigma \Xi$, $f_{S}
( k )^{-1}$ is complex even at the threshold $k = 0 \gev$, which leads
to a complex scattering length.  The real part of $f_{S} ( k )^{-1}$
is negative at the threshold, which implies the existence of an $N
\Omega$ quasibound state below the threshold, and it increases almost
quadratically.  In the same energy region, the imaginary part of
$f_{S} ( k )^{-1}$ almost linearly decreases as a function of $k$ like
$- i k$. Because the energy dependence of $f_{S} ( k )^{-1}$ at low
energy is dictated by $-ik+r_{\rm eff}k^{2}/2$ as shown
Eq.~\eqref{eq:erange_exp}, Fig.~\ref{fig:fSinv} indicates the
imaginary part of the effective range is small.  By using the
formulae~\eqref{eq:scatt_length_full} and \eqref{eq:reff_full}, we can
calculate the scattering length $a$ and effective range $r_{\rm eff}$,
respectively.  In our model we obtain $a = 5.3 - 4.3 i \fm$ and
$r_{\rm eff} = 0.74 + 0.04 i \fm$.  We find that the real part of the
effective range roughly corresponds to the length scale of the
$N\Omega$ interaction, and the imaginary part is small, as
expected. The magnitude of the scattering length is evidently larger
than the interaction range, indicating the $N\Omega$ scattering is
close to the unitary limit. With these threshold parameters, the
effective range expansion~\eqref{eq:erange_exp} reproduces the inverse
of the scattering amplitude $f_{S}^{-1}$ fairly well in the energy
range of Fig.~\ref{fig:fSinv}.

\subsection{\boldmath $N \Omega$ quasibound state}
\label{sec:4D}

Finally, by using the full $N \Omega ( {}^{5} S_{2} )$
interaction~\eqref{eq:Vfull} and solving the Lippmann--Schwinger
equation~\eqref{eq:LSeq_full}, we search for a pole of the $N \Omega (
^{5} S_{2} )$ quasibound state, indicated by the scattering length.
As a result of the numerical calculation, we find a pole of the $N
\Omega ( ^{5} S_{2} )$ quasibound state at $E_{\rm pole} = 2611.3 -
0.7 i \mev$ in the complex energy plane below the $N \Omega$ threshold
(2611.4\mev).  The pole exists in the first Riemann sheet of the $N
\Omega$ channel and in the second Riemann sheets of the $\Lambda \Xi$
and $\Sigma \Xi$ channels.  The pole position corresponds to the
binding energy $0.1 \mev$ and width $1.5 \mev$.

In general, when one takes into account the imaginary part of the
potential to represent absorption into open channels, the binding
energy of a bound state in quantum mechanics decreases. In particular,
a shallow bound state may disappear above the threshold.  In the
present case, the pole position of the $N \Omega ( {}^{5} S_{2} )$
bound state would be $2611.0 \mev$ if the imaginary part of the
interaction coming from the box terms were absent.  The imaginary part
of the $N \Omega$ interaction reduces the binding energy of the $N
\Omega$ bound state from $0.3 \mev$ to $0.1 \mev$.  Therefore, we
confirm that the absorptive effect by the $\Lambda \Xi$ and $\Sigma
\Xi$ channels indeed acts repulsively, but the $N \Omega ( {}^{5}
S_{2} )$ quasibound state stays below the threshold.

The $N\Omega$ system is an isospin doublet, and there are two
components, $p \Omega ^{-}$ and $n \Omega ^{-}$.  In addition to the
strong interaction, in the $p \Omega^{-}$ system, the attractive
Coulomb interaction between $p$ and $\Omega ^{-}$ will assist the
binding more.  This point will be discussed at the end of this
subsection with the wave function of the quasibound state.

In order to investigate the properties of the $N \Omega ( ^{5} S_{2}
)$ quasibound state, we calculate its wave function from the residue
of the $T$-matrix at the pole position, according to the approach in
Ref~\cite{Sekihara:2016xnq}.  The off-shell $T$-matrix contains the
pole in the following expression
\begin{equation}
  T ( E ; \, p^{\prime} , \, p )
  = \frac{\gamma ( p^{\prime} ) \gamma ( p )}{E - E_{\rm pole}}
  + (\text{regular at } E = E_{\rm pole} ) .
  \label{eq:Tpole}
\end{equation}
The function $\gamma ( p )$ is related to the radial part of the
$N \Omega$ quasibound-state wave function in momentum space as:
\begin{equation}
  R_{N \Omega} ( p ) = \frac{\sqrt{4 \pi} \gamma ( p )}
  {E_{\rm pole} - \mathcal{E}_{N \Omega} ( p )} .
\end{equation}
An important point to be noted is that the wave function $R_{N \Omega}
( p )$ is already normalized when extracted from the residue of the
$T$-matrix, as the Lippmann--Schwinger equation~\eqref{eq:LSeq_full}
is inhomogeneous integral equation.

From the wave function $R_{N \Omega} ( p )$ in momentum space we can
calculate the wave function in coordinate space in a straightforward
way:
\begin{align}
  \psi ( r ) = & \int \frac{d^{3} p}{( 2 \pi )^{3}}
  e^{i \bm{p} \cdot \bm{r}} R_{N \Omega} ( p ) Y_{0 0} 
  \notag \\
  = & \frac{Y_{0 0}}{2 \pi ^{2} r} \int _{0}^{\infty} d p \, 
  p \sin ( p r ) R_{N \Omega} ( p ) ,
\end{align}
where $Y_{0 0} \equiv 1 / \sqrt{4 \pi}$ is the spherical harmonics.
The norm of this wave function is expressed as
\begin{align}
  X_{N \Omega} = & \int d^{3} r \,
  \left [ \psi ( r ) \right ] ^{2}
  = \int _{0} ^{\infty} d r \, \text{P}_{N \Omega} ( r ) ,
  \label{eq:norm}
\end{align}
where we define the ``density distribution'' $\text{P}_{N \Omega} ( r
)$:
\begin{align}
  \text{P}_{N \Omega} ( r )
  \equiv 4 \pi r^{2} \left [ \psi ( r ) \right ] ^{2} .
  \label{eq:density}
\end{align}
We note the absence of the complex conjugate in Eqs.~\eqref{eq:Tpole},
\eqref{eq:norm} and \eqref{eq:density}, due to the unstable nature of
the quasibound state. As a consequence, the wave functions $R_{N
  \Omega} ( p )$, $\psi ( r )$, the density distribution $\text{P}_{N
  \Omega} ( r )$, and the norm $X_{N \Omega}$ are in general complex.

The norm $X_{N \Omega}$ from the $T$-matrix is called
compositeness~\cite{Hyodo:2011qc, Hyodo:2013nka, Sekihara:2014kya} and
quantitatively evaluates the importance of the $N \Omega$ degrees of
freedom for the quasibound state in the employed model.  The
compositeness $X_{N \Omega}$ is unity for a purely $N \Omega$ state,
but it deviates from unity when the interaction depends on the energy
$E$ as a consequence of the effective reduction of the inelastic
channels.  In the present formulation, the quasibound state can have
$\Lambda \Xi$, $\Sigma \Xi$, and $\Lambda \Xi ^{\ast}$ components
whose contributions are evaluated with~\cite{Formanek:2003,
  Miyahara:2015bya}
\begin{align}
  X_{j} = & - \frac{1}{16 \pi ^{5}}
  \int _{0}^{\infty} d p \, p^{2} R_{N \Omega} ( p )
  \int _{0}^{\infty} d p^{\prime} \, p^{\prime \, 2} R_{N \Omega} ( p^{\prime} )
  \notag \\
  & \quad \times
  \frac{\partial V_{\text{box}( j )}}{\partial E}
    ( E_{\rm pole} ; \, p^{\prime} , \, p ) .
\end{align}
As a result, we obtain $X_{N \Omega} = 1.00 + 0.00 i$ within three
significant figures while we find that the others $X_{\Lambda \Xi}$,
$X_{\Sigma \Xi}$, and $X_{\Lambda \Xi ^{\ast}}$ are consistent with
zero in this order.  The result indicates that the quasibound state
obtained in the present model is indeed composed of the $N \Omega$
channel.

Besides, using the weak-biding relation derived by
Weinberg~\cite{Weinberg:1965zz}, the compositeness of a shallow bound
state can be determined only by the observable quantities, the
scattering length and the eigenenergy.  The relation extended to the
quasibound state with the finite decay width is given
as~\cite{Kamiya:2015aea, Kamiya:2016oao}
\begin{gather}
  a = R\left[ \frac{2X_{N\Omega}^{\wb}}{1+X_{N\Omega}^{\wb}}+\mathcal{O}\left(\left|
    \frac{R_{\mathrm{typ}}}{R}\right|\right) + \mathcal{O}\left( \left|\frac
         {l}{R}\right|^3 \right) \right],
  \label{eq:WB_rel}
  \\
  R \equiv 1/\sqrt{-2 \mu E_h},\quad  l \equiv 1/ \sqrt{2 \mu \omega},
\end{gather}
where $X_{N \Omega}^{\wb}$ is the compositeness for the $N \Omega$
channel, $E_{h} = E_{\rm pole} - m_{N} - m_{\Omega} = - 0.1 - 0.7 i
\mev$ is the eigenenergy of the bound state measured from the $N
\Omega$ threshold energy, $\mu = m_{N} m_{\Omega} / (m_{N} +
m_{\Omega} )$ is the reduced mass, $R_{\mathrm{typ}}$ is the typical
length scale of the interaction, and $\omega = 37.7\mev$ denotes the
difference between the threshold energy of the $N\Omega$ channel and
that of the nearest channel, $\Lambda\Xi^\ast$.  We estimate
$R_{\mathrm{typ}}$ with the $\eta$ meson exchange interaction as
$R_{\mathrm{typ}} = 1/m_\eta \sim 0.4\fm$. We find that $
|R_{\mathrm{typ}}/R| \sim 0.1$ and $|l/R|^3\sim 0.0$ are much smaller
than unity, which justifies neglecting the second and third terms in
Eq.~\eqref{eq:WB_rel} to calculate the compositeness of the $N\Omega$
quasibound state. Neglecting these correction terms and using the
value of the scattering length $a = 5.3 - 4.3 i \fm$ derived in the
previous subsection, we obtain $X_{N\Omega}^{\wb} = 1.1+0.1i$.  With
this complex $X_{N \Omega}^{\wb}$, the real-valued compositeness,
which is interpreted as the probability~\cite{Kamiya:2015aea,
  Kamiya:2016oao}, is calculated as $\tilde{X}^{\wb}_{N\Omega}
=1.0$. This result indicates the dominance of the $N\Omega$ composite
component, in agreement with the above calculation using the wave
function.

\begin{figure}[!t]
  \centering
  \Psfig{8.6cm}{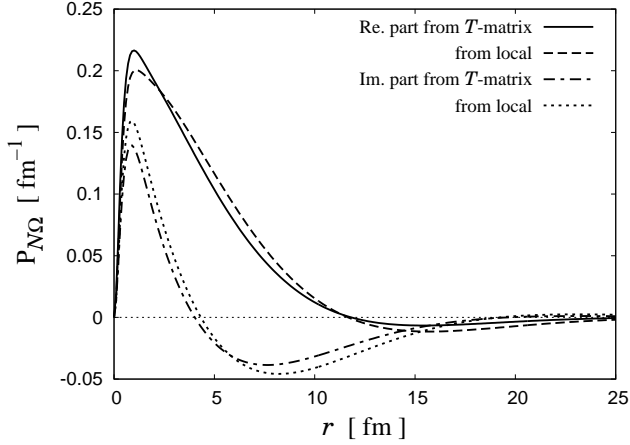}
  \caption{Real and imaginary parts of the density distribution
    $\text{P}_{N \Omega}$ for the $N \Omega ( {}^{5} S_{2} )$
    quasibound state as functions of radial coordinate $r$.  In
    addition to $\text{P}_{N \Omega}$ calculated from the $T$-matrix,
    we also plot $\text{P}_{N \Omega}$ from the equivalent local $N
    \Omega$ potential (see Sec.~\ref{sec:5}). }
  \label{fig:WF}
\end{figure}

We then plot the density distribution $\text{P}_{N \Omega} ( r )$ in
coordinate space~\eqref{eq:density} in Fig.~\ref{fig:WF} (solid and
dash-dotted lines).  Because certain amount of density exists beyond
$r = 10 \fm$, the $N \Omega$ quasibound state is a spatially extended
system owing to the tiny binding energy from the $N \Omega$ threshold.
The average of the ``distance'' between $N \Omega$ is $\sqrt{\langle
  r^{2} \rangle} = 3.8 - 3.1 i \fm$, where we define $\langle r^{2}
\rangle$ as
\begin{align}
  \langle r^{2} \rangle
  \equiv \int _{0}^{\infty} d r \, 
  r^{2} \text{P}_{N \Omega} ( r ) .
  \label{eq:MSR}
\end{align}
Although the ``distance'' is complex due to the resonance nature, its
absolute value largely exceeds the typical size of baryons $0.8 \fm$.
Meanwhile, as explained in Ref.~\cite{Miyahara:2015bya}, the dumping
of the wave function outside the potential range is related to the
standard expectation value of the average distance:
\begin{equation}
  \langle r^{2}_{\rm dump} \rangle \equiv
  \frac{\displaystyle \int _{0}^{\infty} d r \, 
    r^{2} \left | \text{P}_{N \Omega} ( r ) \right |}
       {\displaystyle \int _{0}^{\infty} d r \, 
         \left | \text{P}_{N \Omega} ( r ) \right |} .
       \label{eq:MSR_dump}
\end{equation}
The result is $\sqrt{\langle r^{2}_{\rm dump} \rangle} = 6.5 \fm$,
which indicates dumping of the wave function to a very large distance
compared to the typical hadron scale owing to the tiny binding energy.

We also estimate the shift of the binding energy by the Coulomb
interaction in the $p \Omega ^{-}$ quasibound state.  We calculate the
shift of the bound-state eigenenergy, $\Delta E_{\rm Coulomb}$, in a
perturbation:
\begin{align}
  \Delta E_{\rm Coulomb} 
  \equiv \int _{0}^{\infty} d r \, 
  \left ( - \frac{\alpha}{r} \right ) \text{P}_{N \Omega} ( r ) ,
  \label{eq:dECoulomb}
\end{align}
where $\alpha \approx 1 / 137$ is the fine-structure constant.  The
result is $\Delta E_{\rm Coulomb} = -0.9 - 0.4 i \mev$.  Therefore, we
conclude that both the binding energy and decay width will
respectively shift $\sim + 1 \mev$ by the Coulomb interaction for the
$p \Omega ^{-}$ bound state.

\begin{table*}[t]
  \caption{Parameters $C_{n}$ for an equivalent local $N \Omega$
    interaction.  Other quantities in Eq.~\eqref{eq:Vlocal} are fixed
    as $\Lambda = 1 \gev$ and $m_{n} = n \times ( 100 \mev )$.}
  \label{tab:para}
  \begin{ruledtabular}
    \begin{tabular}{crrrrrrr}
      $n$ & A & B & C &
      box $\Lambda \Xi ( {}^{3} D_{2} + {}^{1} D_{2} ) $ &
      box $\Sigma \Xi ( {}^{3} D_{2} + {}^{1} D_{2} ) $ &
      box $\Lambda \Xi ^{\ast}$ &
      Total
      \\
      \hline
      1 &
      $0.02$ &
      $0.06$ &
      $0.07$ &     
      $-0.04 + \phantom{00} 0.00 i$ &
      $-0.01 + \phantom{00} 0.00 i$ &
      $0.04$ &
      $0.14 + \phantom{00} 0.00 i$
      \\
      2 & 
      $-2.37$ &
      $-6.21$ &
      $-6.48$ &
      $4.67 - \phantom{00} 0.19 i$ &
      $0.93 + \phantom{00} 0.05 i$ &
      $-4.30$ &
      $-13.76 - \phantom{00} 0.14 i$
      \\
      3 &
      $57.03$ &
      $160.19$ &
      $131.39$ &
      $-121.94 + \phantom{00} 5.34 i$ &
      $-24.01 - \phantom{00} 1.24 i$ &
      $104.15$ &
      $306.81 + \phantom{00} 4.10 i$
      \\
      4 &
      $-556.75$ &
      $-1680.33$ &
      $-1021.60$ &
      $1304.16 - \phantom{0} 59.70 i$ &
      $251.48 + \phantom{0} 12.83 i$ &
      $-1026.45$ &
      $-2729.49 - \phantom{0} 46.87 i$
      \\
      5 &
      $2699.73$ &
      $8765.95$ &
      $3548.93$ &
      $-6980.93 + 287.70 i$ &
      $-1313.42 - \phantom{0} 62.87 i$ &
      $5024.22$ &
      $11744.48 + 224.83 i$
      \\
      6 &
      $-7052.95$ &
      $-24755.80$ &
      $-5159.25$ &
      $20223.50 -534.01 i$ &
      $3719.98 + 167.01 i$ &
      $-13263.90$ &
      $-26288.42 - 367.01 i$
      \\
      7 &
      $10055.50$ &
      $38369.80$ &
      $667.40$ &
      $-31881.20 + \phantom{0} 69.44 i$ &
      $-5772.02 -262.54 i$ &
      $19118.60$ &
      $30558.08 - 193.09 i$
      \\
      8 &
      $-7304.99$ &
      $-30596.40$ &
      $5175.64$ &
      $25509.10 + 685.14 i$ &
      $4577.55 + 227.55 i$ &
      $-14091.70$ &
      $-16730.80 + 912.69 i$
      \\
      9 &
      $2096.47$ &
      $9776.40$ &
      $-3446.89$ &
      $-8069.25 - 460.13 i$ &
      $-1442.98 - \phantom{0} 81.38 i$ &
      $4138.10$ &
      $3051.85 - 541.51 i$
    \end{tabular}
  \end{ruledtabular}
\end{table*}

\section{\boldmath Equivalent local $N \Omega$ potential}
\label{sec:5}

The existence of the $N\Omega$ quasibound state below the threshold
implies possible $\Omega$ nuclei, generated by the attractive $N
\Omega$ interaction.  In addition, such possible $\Omega$ nuclei would
shift eigenenergies of $\Omega ^{-}$ atoms, Coulombic bound states of
$\Omega ^{-}$ and nuclei.  To study the few-body system of $\Omega$ in
nuclei, it is useful to have a local $N\Omega$ potential in the
Schr\"odinger equation for which several established techniques to
perform rigorous few-body calculations are available. On the other
hand, because the momentum-space $N \Omega$ interaction in the present
formulation~\eqref{eq:Vfull} is a function not only of the momentum
transfer $| \bm{p} - \bm{p}^{\prime} |$ but also of the momenta $p$
and $p^{\prime}$ individually, it is in general nonlocal.  In
addition, the scattering equation~\eqref{eq:LSeq_full} is formulated
with semirelativistic kinematics for baryons. Here we construct a
local potential which equivalently reproduce the $N \Omega ( {}^{5}
S_{2} )$ scattering amplitude in this study. We first determine the
local potential through the matching with the
interaction~\eqref{eq:Vfull}, and then check whether the low-energy
observables are properly reproduced.

We consider a local potential in the $S$-wave Schr\"odinger equation
\begin{equation}
  \left[
    -\frac{1}{2 \mu r} \frac{d^{2}}{d r^{2}} r
    + m_{N} + m_{\Omega} + V_{\rm local}(r)
  \right] \psi(r)
  = E \psi(r),
  \label{eq:Schroedinger}
\end{equation}
where $r$ is the relative coordinate of the $N\Omega$ system and the
reduced mass is defined as $\mu = m_{N} m_{\Omega} / (m_{N} +
m_{\Omega} )$. Note that the mass energy is included in the
Hamiltonian, in order to be consistent with the definition of $E$ in
this paper. To parametrize $V_{\rm local}(r)$, we introduce an
analytic potential in momentum space as a superposition of nine Yukawa
terms with different exchanged mass $m_{n}$:
\begin{equation}
  \tilde{V}_{\rm local} ( q )
  = \sum _{n = 1}^{9} \frac{C_{n}}{q^{2} + m_{n}^{2}} 
  \left ( \frac{\Lambda ^{2}}{\Lambda ^{2} + q^{2}} \right ) ^{2} ,
  \label{eq:Vlocal}
\end{equation}
where $q$ is the momentum transfer, $\Lambda$ is a cutoff, and $C_{n}$
are the strength parameters of the local potential.  This local
potential in coordinate space is expressed as
\begin{align}
  V_{\rm local} ( r ) =
  & 
  \int \frac{d^{3} q}{( 2 \pi )^{3}}
  e^{- i \bm{q} \cdot \bm{r}} \tilde{V}_{\rm local} ( q )
  \notag \\ 
  = & \frac{1}{4 \pi r} \sum _{n=1}^{9} C_{n}
  \left ( \frac{\Lambda ^{2}}{\Lambda ^{2} - m_{n}^{2}} \right ) ^{2}
  \notag \\ & \times
  \left [ e^{- m_{n} r}
    - \frac{(\Lambda ^{2} - m_{n}^{2}) r + 2 \Lambda}{2 \Lambda}
    e^{- \Lambda r} \right ] .
\end{align}
To determine the strength parameters $C_{n}$, we project the
momentum-space potential to the $S$ wave as
\begin{align}
  & V_{\rm NR} ( p^{\prime} , \, p )
  \notag \\ &
  = \frac{1}{2} \int _{-1}^{1} d \cos \theta \,
  \tilde{V}_{\rm local}
  \left ( \sqrt{p^{2} - 2 p p^{\prime} \cos \theta + p^{\prime \, 2}} \right )
  \notag \\ &
  = \frac{1}{4 p p^{\prime}} \sum _{n=1}^{9} C_{n} 
  \left ( \frac{\Lambda ^{2}}{\Lambda ^{2} - m_{n}^{2}} \right ) ^{2}
  \notag \\ & \quad \times
  \left \{ \log \left [ \frac{( p + p^{\prime} )^{2} + m_{n}^{2}}
    {( p - p^{\prime} )^{2} + m_{n}^{2}} \right ]
  - \log \left [ \frac{( p + p^{\prime} )^{2} + \Lambda ^{2}}
    {( p - p^{\prime} )^{2} + \Lambda ^{2}} \right ] \right .
  \notag \\ & \quad \quad
  \left .
  + \frac{\Lambda ^{2} - m_{n}^{2}}{( p + p^{\prime} )^{2} + \Lambda ^{2}}
  - \frac{\Lambda ^{2} - m_{n}^{2}}{( p - p^{\prime} )^{2} + \Lambda ^{2}}
  \right \} .
\end{align}
The $S$-wave Lippmann--Schwinger equation to obtain the T-matrix
$T_{\rm NR}$, corresponding to the Schr\"odinger
equation~\eqref{eq:Schroedinger}, is expressed as
\begin{align}
  & T_{\rm NR} ( E ; \, p^{\prime} , \, p )
  = V_{\rm NR} ( p^{\prime} , \, p )
  \notag \\
  & \quad
  + \int _{0}^{\infty}
  \frac{d p^{\prime \prime}}{2 \pi ^{2}} p^{\prime \prime \, 2}
  \frac{V_{\rm NR} ( p^{\prime} , \, p^{\prime \prime} )
    T_{\rm NR} ( E ; \, p^{\prime \prime} , \, p )}
       {E - \mathcal{E}_{\rm NR} ( p^{\prime \prime} ) + i 0} ,
       \label{eq:LSeq_NR}
\end{align}
where the nonrelativistic on-shell energy is $\mathcal{E}_{\rm NR} ( p
) \equiv m_{N} + m_{\Omega} + p^{2} / ( 2 \mu )$.  We determine
$C_{n}$ by the matching of $V_{\rm NR} ( p^{\prime} , \, p )$ with
$V(E ; \, p^{\prime} , \, p)$ in Eq.~\eqref{eq:Vfull} at the threshold
energy $E = m_{N} + m_{\Omega}$ as:
\begin{equation}
  V (E= m_{N} + m_{\Omega};\, p^{\prime} , \, p )=
  f ( p ) f ( p^{\prime} )  V_{\rm NR} ( p^{\prime} , \, p ) ,
  \label{eq:matching}
\end{equation}
with a factor to compensate the difference of the kinematics
\begin{equation}
  f ( p ) \equiv \sqrt{\frac{\mathcal{E}_{N \Omega} ( p ) - m_{N} - m_{\Omega}}
    {p^{2} / ( 2 \mu )}} .
\end{equation}
With the factors $f ( p ) f ( p^{\prime} )$, Eq.~\eqref{eq:LSeq_NR}
coincides with the Lippmann--Schwinger equation~\eqref{eq:LSeq_full}
at the threshold.

We set the cutoff as the same value with $V(E ; \, p^{\prime} , \, p)$
in Eq.~\eqref{eq:Vfull}, $\Lambda = 1 \gev$, and the mass parameters
are chosen to be $m_{n} = n \times ( 100 \mev )$ to cover the relevant
ranges of the $N\Omega$ interaction.  Then, we fit the coefficients
$C_{n}$ to satisfy the condition~\eqref{eq:matching}.  With nine terms
in Eq.~\eqref{eq:Vlocal}, we can reproduce each component of $N \Omega
( {}^{5} S_{2} )$ interaction in Eq.~\eqref{eq:Vfull} fairly well in
the whole $p$-$p^{\prime}$ plane.  As a result of the best fit, we
obtain the parameters $C_{n}$ listed in Table~\ref{tab:para}.

Now we check that the local potential $V_{\rm local} ( r )$ well
reproduces properties of the $N \Omega ( {}^{5} S_{2} )$ scattering
amplitude around the threshold energy. Because we neglect the energy
dependence of the potential, the local potential cannot be
extrapolated to the energy region far away from the threshold. In the
following, we examine the eigenenergy of the $N\Omega$ quasibound
state and the low-energy scattering with momentum $k\leq 0.2$ GeV.

First, we solve the \Schr equation~\eqref{eq:Schroedinger} with the
local potential and obtain a quasibound state with eigenenergy $E =
2611.4 - 0.7 i \mev$, which reproduces the pole position of the
$T$-matrix in Eq.~\eqref{eq:LSeq_full} to an accuracy of $0.1 \mev$.
From the wave function of the quasibound state, we calculate the
density distribution $\text{P}_{N \Omega}$ as in
Eq.~\eqref{eq:density} and normalize it by the condition $X_{N \Omega}
= 1$ in Eq.~\eqref{eq:norm}.  The real and imaginary parts of the
resulting density distribution are plotted in Fig.~\ref{fig:WF} by the
dashed and dotted lines, respectively.  We can see that the density
distribution from the local potential is very similar to that from the
$T$-matrix.  We also calculate the ``distance'' between $N \Omega$,
which results in $\sqrt{\langle r^{2} \rangle} = 2.8 - 4.5 i \fm$ and
$\sqrt{\langle r^{2}_{\rm dump} \rangle} = 7.4 \fm$ in the
prescriptions of Eq.~\eqref{eq:MSR} and Eq.~\eqref{eq:MSR_dump},
respectively.  These values are in fair agreement with those from the
$T$-matrix as well.

Let us switch on the Coulomb potential $V_{\rm Coulomb} ( r ) = -
\alpha / r$ for the $p \Omega ^{-}$ system.  In the calculation of the
energy shift in a perturbation of Eq.~\eqref{eq:dECoulomb}, we would
obtain a similar result as in the previous section, because of the
similarity of the density distributions $\text{P}_{N \Omega}$.
Instead of such a perturbative calculation, we can easily perform the
full calculation in the present case by solving the \Schr
equation~\eqref{eq:Schroedinger} with $V_{\rm Coulomb} + V_{\rm
  local}$.  As a result of the full calculation, the eigenenergy moves
to $2610.5 - 1.0 i \mev$, where the binding energy and decay width
shift $+ 0.9 \mev$ and $+ 0.6 \mev$, respectively.  The increase of
the binding energy is a natural consequence of the attractive Coulomb
interaction.  The Coulomb attraction induces the shrinkage of the wave
function of the $N \Omega$ system, which leads to the increase of the
decay width due to the enlarged overlap of two particles.  The result
of the shift of the eigenenergy indicates that the perturbative
calculation of Eq.~\eqref{eq:dECoulomb} gives a good estimation.

Second, we calculate the $S$-wave scattering amplitude $f_{S} ( k )$
from the asymptotic behavior of the wave function $\psi(r)$ at energy
$E$ with the local potential $V_{\rm local} ( r )$.  The resulting
$f_{S} ( k )^{-1}$ is plotted in Fig.~\ref{fig:fSinv} as the dashed
and dotted lines.  We find that $f_{S} ( k )^{-1}$ nicely reproduces
the result from the $T$-matrix, which means that the local potential
$V_{\rm local}$ is accurate enough to describe the $N \Omega ( {}^{5}
S_{2} )$ scattering near the threshold $k\leq 0.2$ GeV.  With $f_{S} (
k )^{-1}$ from the local potential $V_{\rm local} ( r )$, we evaluate
the scattering length and effective range as $a = 5.2 - 5.0 i \fm$ and
$r_{\rm eff} = 0.78 + 0.06 i \fm$, in fair agreement with those from
the $T$-matrix.  Note that the value of the scattering length is
sensitive to the small modification of the system, reflecting the
divergence in the unitary limit.

\begin{figure}[!t]
  \centering
  \Psfig{8.6cm}{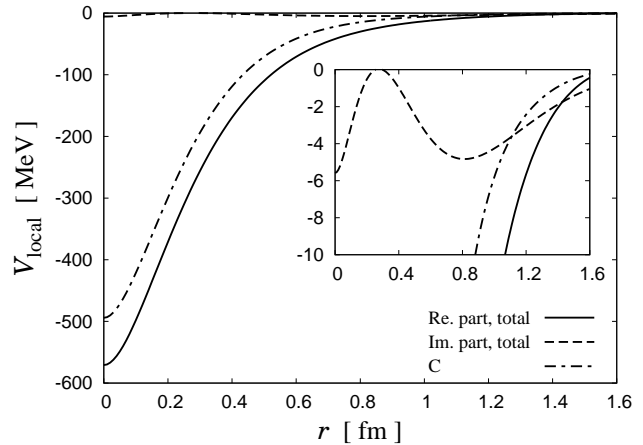}
  \caption{Equivalent local $N \Omega$ potential in coordinate space
    $V_{\rm local} (r)$. The contribution from the contact term
    $V_{\rm C}$ is also shown for comparison. The inset represents an
    enlarged figure.}
  \label{fig:Vr}
\end{figure}

We show in Fig.~\ref{fig:Vr} the equivalent local $N \Omega ( {}^{5}
S_{2} )$ potential in coordinate space $V_{\rm local} (r)$ together
with the contribution from the contact term $V_{\rm C}$.  From the
figure, we confirm that the strong attraction in the $N \Omega (
{}^{5} S_{2} )$ interaction originates from the contact term $V_{\rm
  C}$ while other contributions give moderate attraction.  The
interaction range in Fig.~\ref{fig:Vr} is consistent with the
effective range $\sim 0.7 \fm$ obtained from the scattering amplitude
$f_{S}$.

\section{Conclusion}
\label{sec:6}

In this study we have investigated the $N \Omega ( {}^{5}S_{2} )$
interaction based on a baryon--baryon interaction model with meson
exchanges.  The long-range part has been composed of the conventional
mechanisms: exchanges of $\eta$ and ``$\sigma$'', i.e., correlated two
mesons in the scalar-isoscalar channel.  The short-range part has been
represented by the contact interaction.  In addition, we have taken
into account inelastic channels $\Lambda \Xi$, $\Sigma \Xi$, and
$\Lambda \Xi (1530)$ which couple to the $N \Omega ( {}^{5} S_{2} )$
system via $K$ exchange.  The inclusion of the open channels, $\Lambda
\Xi$ and $\Sigma \Xi$, is important to describe the absorption effects
in the physical $N\Omega$ system. The unknown strength of the contact
interaction was determined by fitting the scattering length of the HAL
QCD result at the nearly physical quark masses.

The constructed $N\Omega ({}^{5} S_{2} )$ interaction was used to
calculate the observable quantities at the physical point, including
the absorption effects.  For the $N \Omega ({}^{5} S_{2} )$
scattering, we have obtained the scattering length $a = 5.3 - 4.3 i
\fm$ and the effective range $r_{\rm eff} = 0.74 + 0.04 i \fm$. The
larger magnitude of the scattering length than the effective range
indicates that the $N\Omega$ interaction is close to the unitary
limit, and the positive real part indicates the existence of a shallow
quasibound state below the threshold.  Indeed, searching for the pole
of the scattering amplitude, we have found that the $N \Omega ( {}^{5}
S_{2} )$ quasibound state is generated with its eigenenergy $2611.3 -
0.7 i \mev$, which corresponds to the binding energy $0.1 \mev$ and
the width $1.5 \mev$.  When the imaginary part of the interaction is
switched off, we obtain a bound state at 2611.0 MeV. Thus, the
imaginary part primarily induces the decay width, and slightly reduces
the binding energy.  The main decay mode is $\Lambda \Xi$, owing to
the larger $K N \Lambda$ coupling than the $KN\Sigma$ one.  For the $p
\Omega ^{-}$ bound state, the attractive Coulomb interaction further
add a shift of $\sim + 1 \mev$ both to the binding energy and decay
width.  The spatial size of the $N \Omega$ bound state will largely
exceed the typical size of baryons.

We have discussed how the different mechanisms contribute to the
$N\Omega$ interaction. It turns out that the attraction dominantly
originates from the contact term.  Other contributions, the $\eta$
exchange, correlated two-meson exchange, and box terms with inelastic
channels in the intermediate states, give moderate attraction.
Because we have considered all conventional mechanisms at the hadronic
level, the discussion at the quark-gluon level would be necessary to
clarify the origin of the $N \Omega$ attraction.  Although the
elimination of the inelastic channels induces the energy dependence of
the single-channel $N \Omega$ interaction, the energy dependence has
been found to be less than 1 \% in the energy region 50 MeV above and
below the threshold.  We have found that the contribution from the
$D$-wave $N \Omega$ states to the $N \Omega ({}^{5} S_{2} )$
interaction is negligible as well.  These results justify constructing
a single-channel $N\Omega$ potential in $S$ wave in the HAL QCD
analysis~\cite{Doi:2017zov, Iritani:2018a}.

We have constructed an equivalent local $N \Omega ({}^{5} S_{2} )$
potential, which will be useful to applications for few-body systems,
such as possible $\Omega$ nuclei generated by the attractive $N
\Omega$ interaction.  To avoid the $S$-wave decays which would bring a
large decay width, it is essential to align the spins of $\Omega$ and
nucleons to the same direction so that the $N \Omega$ system couples
to the $\Lambda \Xi$ and $\Sigma \Xi$ decay modes only in $D$ wave.
In this sense, the $\Omega$-deuteron bound state with $J^{P} =
5/2^{+}$ will be the most plausible candidate of the $\Omega$ nuclei.

Finally, we remark the possibility of the experimental investigation
of the $N\Omega$ interaction and the quasibound state. Because the
$N\Omega$ system has strangeness $S=-3$, practical candidate is the
production in heavy ion collisions~\cite{Cho:2010db,
  Cho:2011ew}. Thanks to the small decay width, the $N\Omega$
quasibound state should be observed as a narrow peak in the invariant
mass spectrum of the $\Lambda \Xi$ system near the $N\Omega$
threshold. In fact, the production yield of the $N\Omega$ bound state
is estimated in Ref.~\cite{Cho:2017dcy} to be of the order of
$10^{-3}$ per central collision at RHIC and LHC, assuming the binding
energy of the $N\Omega$ system as 19 MeV. If the binding energy is
much smaller as we find in this paper, the production yield should be
enhanced. Another tool is, as discussed in Ref.~\cite{Morita:2016auo},
the $p\Omega^{-}$ two-body correlation which reflects the low-energy
interaction of the $p\Omega^{-}$ system. Because we have shown that
the imaginary part of the scattering length has the same magnitude
with the real part, the coupling to the open channels should be taken
into account to study the realistic $p\Omega^{-}$ correlation
function.

\begin{acknowledgments}
  The authors acknowledge A.~Ohnishi, F.-K.~Guo, and E.~Oset for
  fruitful discussions on the recent studies of the $N \Omega$ bound
  state, M.~Oka on the dibaryons, and H.~Kamano on the partial wave
  analysis with helicity eigenstates.  The authors are grateful to
  T.~Doi, T.~Iritani, and S.~Gongyo for helpful discussions on the HAL
  QCD potential for the $N \Omega$ system.  T.~S.\ would like to thank
  the Yukawa Institute for Theoretical Physics, Kyoto University,
  where this work was initiated, for kind hospitality during his
  visit.

  This work was partly supported by JSPS KAKENHI Grants
  No.~JP15K17649, No.~JP16K17694, No.~JP17J04333, by JSPS Japan-France
  Joint Research Project, and by the Yukawa International Program for
  Quark-Hadron Sciences (YIPQS).

\end{acknowledgments}

\appendix

\section{Masses and widths of hadrons}
\label{app:masses}

In this study we use isospin symmetric masses for
hadrons~\cite{Olive:2016xmw}: $m_{\pi} = 138.0 \mev$, $m_{K} = 495.6
\mev$ and $m_{\eta} = 547.9 \mev$ for mesons, $m_{N} = 938.9 \mev$,
$m_{\Lambda} = 1115.7 \mev$, $m_{\Sigma} = 1193.2 \mev$, and $m_{\Xi}
= 1318.3 \mev$, for octet baryons, and $m_{\Delta} = 1210.0 \mev$,
$m_{\Sigma ^{\ast}} = 1384.6 \mev$, $m_{\Xi ^{\ast}} = 1533.4 \mev$,
and $m_{\Omega} = 1672.5 \mev$ for decuplet baryons.  In addition, the
widths of the decuplet baryons are $\Gamma _{\Delta} = 100.0 \mev$,
$\Gamma _{\Sigma ^{\ast}} = 37.1 \mev$, and $\Gamma _{\Xi ^{\ast}} =
9.5 \mev$.

\section{Conventions}
\label{app:conv}

In this Appendix we summarize our conventions of baryons used in this
study.

Throughout this study the metric in four-dimensional Minkowski space
is $g^{\mu \nu} = g_{\mu \nu} = \text{diag} ( 1 , \, -1 , \, -1 , \,
-1 )$ and the Einstein summation convention is used.  The Dirac
matrices $\gamma ^{\mu}$ satisfy
\begin{equation}
  \{ \gamma ^{\mu} , \, \gamma ^{\nu} \} = 2 g^{\mu \nu} .
\end{equation}
In the present study we choose the standard representation for the
Dirac matrices:
\begin{equation}
  \gamma ^{0} =
  \left(
  \begin{array}{@{\,}cc@{\,}}
    \bm{1} & 0 \\
    0 & - \bm{1} 
  \end{array}
  \right)
  , \quad 
  \bm{\gamma} = \left(
  \begin{array}{@{\,}cc@{\,}}
    0 & \bm{\sigma} \\
    - \bm{\sigma} & 0 
  \end{array}
  \right) ,
\end{equation}
with the Pauli matrices $\bm{\sigma}$, and
\begin{equation}
  \gamma ^{5} \equiv i \gamma ^{0} \gamma ^{1} 
  \gamma ^{2} \gamma ^{3} 
  = \left(
  \begin{array}{@{\,}cc@{\,}}
    0 & \bm{1} \\
    \bm{1} & 0 
  \end{array}
  \right) .
\end{equation}

The Dirac spinors for a positive energy solution are expressed as $u (
\bm{p} , \, s )$ with its three-momentum $\bm{p}$ and helicity
$\lambda$, and its normalization is
\begin{equation}
  \bar{u} ( \bm{p} , \, \lambda ^{\prime} ) u ( \bm{p} , \, \lambda  ) =
  \delta _{\lambda ^{\prime} \lambda} ,
\end{equation}
where $\bar{u} \equiv u^{\dagger} \gamma ^{0}$.  We employ the
following explicit form of the Dirac spinors
\begin{equation}
  u ( \bm{p} , \, \lambda  )
  = \sqrt{\frac{E ( p ) + M}{2 M}} \left (
  \begin{array}{@{\,}c@{\,}}
    \phantom{\displaystyle \frac{p}{E_{p}}} \chi _{\lambda}
    \phantom{\displaystyle \frac{p}{E_{p}}} \\
    \displaystyle \frac{\bm{\sigma} \cdot \bm{p}}{E ( p ) + M} \chi _{\lambda}
  \end{array}
  \right ) ,
\label{eq:u_spinor}
\end{equation}
where $M$ is the mass of the particle, $p \equiv |\bm{p}|$, and $E ( p
) \equiv \sqrt{p^{2} + M^{2}}$.  The two-component spinor $\chi
_{\lambda}$ is chosen to be helicity eigenstates
\begin{equation}
  \begin{split}
    & \chi _{+1/2} = \left (
    \begin{array}{@{\,}c@{\,}}
      e^{- i \phi / 2} \cos ( \theta / 2 ) \\
      e^{+ i \phi / 2} \sin ( \theta / 2 ) 
    \end{array}
    \right ) ,
    \\
    & \chi _{-1/2} = \left (
    \begin{array}{@{\,}c@{\,}}
      - e^{- i \phi / 2} \sin ( \theta / 2 ) \\
      e^{+ i \phi / 2} \cos ( \theta / 2 ) 
    \end{array}
    \right ) ,
  \end{split}
\end{equation}
where $\theta$ and $\phi$ are the polar and azimuthal angles of the
momentum $\bm{p}$, respectively.  In fact $\chi _{\lambda}$ satisfies
\begin{equation}
  \frac{1}{2} \frac{\bm{\sigma} \cdot \bm{p}}{p}
  \chi _{\pm 1/2} = \pm \frac{1}{2}
  \chi _{\pm 1/2} .
\end{equation}
Owing to this relation, one can simplify the spinors in
Eq.~\eqref{eq:u_spinor} as
\begin{equation}
  u ( \bm{p} , \, \lambda  )
  = \left (
  \begin{array}{@{\,}c@{\,}}
    \displaystyle \sqrt{\frac{E ( p ) + M}{2 M}} \, \chi _{\lambda} 
    \vspace{5pt} \\
    \displaystyle \sqrt{\frac{E ( p ) - M}{2 M}} \, 2 \lambda \chi _{\lambda}
  \end{array}
  \right ) .
\label{eq:u_spinor-II}
\end{equation}
The Dirac spinors for a negative energy solution is then calculated as
$v ( \bm{p} , \, \lambda ) \equiv i \gamma ^{2} u^{\ast} ( \bm{p} , \,
\lambda )$, or explicitly
\begin{equation}
  v ( \bm{p} , \, \lambda  )
  = \left (
  \begin{array}{@{\,}c@{\,}}
    \displaystyle - \sqrt{\frac{E ( p ) - M}{2 M}} \, \chi _{- \lambda} 
    \vspace{5pt} \\
    \displaystyle \sqrt{\frac{E ( p ) + M}{2 M}} \, 2 \lambda \chi _{- \lambda}
  \end{array}
  \right ) ,
\end{equation}
which is normalized as 
\begin{equation}
  \bar{v} ( \bm{p} , \, \lambda ^{\prime} ) v ( \bm{p} , \, \lambda  ) =
  - \delta _{\lambda ^{\prime} \lambda} .
\end{equation}
The Dirac spinors satisfy the following relations
\begin{equation}
  ( \Slash{p} - M ) u ( \bm{p} , \, \lambda ) = 0 ,
  \quad 
  ( \Slash{p} + M ) v ( \bm{p} , \, \lambda ) = 0 ,
\end{equation}
and
\begin{equation}
  \begin{split}
    & \sum _{\lambda} u ( \bm{p} , \, \lambda ) \bar{u} ( \bm{p} , \, \lambda )
    = \frac{\Slash{p} + M}{2 M} ,
    \\
    & \sum _{\lambda} v ( \bm{p} , \, \lambda ) \bar{v} ( \bm{p} , \, \lambda )
    = \frac{\Slash{p} - M}{2 M} ,
  \end{split}
\end{equation}
with $p^{\mu} = ( E ( p ), \, \bm{p})$ and $\Slash{p} \equiv p_{\mu}
\gamma ^{\mu}$.

Next, $e^{\mu} ( \bm{p} , \, \lambda )$ denotes the polarization
vectors for the spin $1$ particle of three-momentum $\bm{p}$, helicity
$\lambda$, and mass $M$.  The vectors are chosen to be helicity
eigenstates:
\begin{equation}
  \begin{split}
    & e^{\mu} ( \bm{p} , \, \pm 1 )
    =
    \frac{\pm 1}{\sqrt{2}} ( 0, \, - \cos \phi \cos \theta \pm i \sin \phi , \,
    \\
    & 
    \phantom{e^{\mu} ( \bm{p} , \, \pm 1 )
    =
    \sqrt{2} ( 0, \, - }
    - \sin \phi \cos \theta \mp i \cos \phi , \, \sin \theta ) ,
    \\
    & e^{\mu} ( \bm{p} , \, 0 )
    = \frac{E ( p )}{M} \left ( \frac{p}{E ( p )} , \,
    \cos \phi \sin \theta , \,
    \sin \phi \sin \theta , \, \cos \theta \right ) .
  \end{split}
  \label{eq:e_spinor}
\end{equation}
These are obtained by boosting the helicity eigenstates at the rest
frame of the particle $e^{\mu} ( \bm{0} , \, \lambda ) = ( 0, \,
\hat{e}_{\lambda} )$ with the three-vector $\hat{e}_{\lambda}$
\begin{equation}
  \begin{split}
    & \hat{e}_{+1} = \left ( - \frac{1}{\sqrt{2}} , \, - \frac{i}{\sqrt{2}} ,
    \, 0 \right ) ,
    \quad \hat{e}_{0} = ( 0 , \, 0 , \, 1 ) ,
    \\
    & \hat{e}_{-1} = \left ( \frac{1}{\sqrt{2}} , \, - \frac{i}{\sqrt{2}} ,
    \, 0 \right ) ,
  \end{split}
\end{equation}
to the direction of the $z$ axis to have a momentum $p$ and then
rotate to the direction of $( \theta , \, \phi)$, where the momentum
is $\bm{p} = (p \cos \phi \sin \theta , \, p \sin \phi \sin \theta ,
\, p \cos \theta )$.  The polarization vectors are normalized as
\begin{equation}
  e^{\mu} ( \bm{p} , \, \lambda ^{\prime} )
  e_{\mu}^{\ast} ( \bm{p} , \, \lambda ) = - \delta _{\lambda ^{\prime} \lambda } ,
\end{equation}
and satisfy the following relations
\begin{equation}
  p^{\mu} e_{\mu} ( \bm{p} , \, \lambda ) = 0 ,
\end{equation}
and
\begin{equation}
  \sum _{\lambda} e^{\mu} ( \bm{p} , \, \lambda )
  e^{\ast \, \nu} ( \bm{p} , \, \lambda )
  = - g^{\mu \nu} + \frac{p^{\mu} p^{\nu}}{M^{2}} .
\end{equation}

Finally, the Rarita--Schwinger spinors for the spin $3/2$ particle,
$u^{\mu} ( \bm{p} , \, \lambda )$, are constructed from the Dirac
spinors $u ( \bm{p} , \, \lambda )$ in Eq.~\eqref{eq:u_spinor-II} and
polarization vectors $e^{\mu} ( \bm{p} , \, \lambda )$ in
Eq.~\eqref{eq:e_spinor} as
\begin{equation}
  u^{\mu} ( \bm{p} , \, \lambda )
  = \sum _{\lambda _{1} , \lambda _{2}}
  \langle 1 \, 1/2 \, \lambda _{1} \, \lambda _{2} | 3/2 \, \lambda \rangle
  e^{\mu} ( \bm{p} , \, \lambda _{1} ) u ( \bm{p} , \, \lambda _{2} ) ,
\end{equation}
with the Clebsch--Gordan coefficients $\langle j_{1} \, j_{2} \, m_{1}
\, m_{2} | J \, M \rangle$.  More explicitly, the Rarita--Schwinger
spinors are
\begin{equation}
  \begin{split}
    & u^{\mu} ( 3 / 2 )
    = e^{\mu} ( 1 ) u ( 1 / 2 ) ,
    \\
    & u^{\mu} ( 1 / 2 )
    = \sqrt{\frac{2}{3}} e^{\mu} ( 0 ) u ( 1 / 2 )
    + \sqrt{\frac{1}{3}} e^{\mu} ( 1 ) u ( -1 / 2 ) ,
    \\
    & u^{\mu} ( -1 / 2 )
    = \sqrt{\frac{1}{3}} e^{\mu} ( -1 ) u ( 1 / 2 ) 
    + \sqrt{\frac{2}{3}} e^{\mu} ( 0 ) u ( -1 / 2 ) ,
    \\
    & u^{\mu} ( - 3 / 2 )
    = e^{\mu} ( - 1 ) u ( - 1 / 2 ) ,
  \end{split}
\end{equation}
where we omitted the argument $\bm{p}$ for the Dirac spinors and
polarization vectors.  The Rarita--Schwinger spinors are normalized as
\begin{equation}
  \bar{u}^{\mu} ( \bm{p} , \, \lambda ^{\prime} )
  u_{\mu} ( \bm{p} , \, \lambda ) = - \delta _{\lambda ^{\prime} \lambda } ,
\end{equation}
and satisfy the following relations
\begin{equation}
  ( \Slash{p} - M ) u_{\mu} ( \bm{p} , \, \lambda ) = 0 ,
\end{equation}
\begin{equation}
  p^{\mu} u_{\mu} ( \bm{p} , \, \lambda ) = 0 ,
  \quad 
  \gamma ^{\mu} u_{\mu} ( \bm{p} , \, \lambda ) = 0 .
\end{equation}
In addition, at the rest frame of the particle, the Rarita--Schwinger
spinors satisfy $u^{0} ( \bm{0} , \, \lambda ) = 0$ and
\begin{equation}
  \sum _{\lambda} u^{i} ( \bm{0} , \, \lambda )
  \bar{u}^{j} ( \bm{0} , \, \lambda )
  = \left(
  \begin{array}{@{\,}cc@{\,}}
    2 \delta _{i j} / 3 - i \epsilon _{i j k} \sigma ^{k} / 3 & 0 \\
    0 & 0 
  \end{array}
  \right) .
\end{equation}
In a similar manner, we can construct the Rarita--Schwinger spinors
for the antiparticle, $v^{\mu} ( \bm{p} , \, \lambda )$, as
\begin{equation}
  v^{\mu} ( \bm{p} , \, \lambda )
  = \sum _{\lambda _{1} , \lambda _{2}}
  \langle 1 \, 1/2 \, \lambda _{1} \, \lambda _{2} | 3/2 \, \lambda \rangle
  e^{\ast \, \mu} ( \bm{p} , \, \lambda _{1} ) v ( \bm{p} , \, \lambda _{2} ) ,
\end{equation}
which satisfy the following normalization and relations
\begin{equation}
  \bar{v}^{\mu} ( \bm{p} , \, \lambda ^{\prime} )
  v_{\mu} ( \bm{p} , \, \lambda ) = \delta _{\lambda ^{\prime} \lambda } ,
\end{equation}
\begin{equation}
  ( \Slash{p} + M ) v_{\mu} ( \bm{p} , \, \lambda ) = 0 ,
\end{equation}
\begin{equation}
  p^{\mu} v_{\mu} ( \bm{p} , \, \lambda ) = 0 ,
  \quad 
  \gamma ^{\mu} v_{\mu} ( \bm{p} , \, \lambda ) = 0 .
\end{equation}

\section{Partial-wave projection of interactions}
\label{app:proj}

In this Appendix we show formulae of the projection of baryon--baryon
interactions to general partial waves.  Here the baryon--baryon
scatterings are denoted by $B_{1} ( p_{1}^{\mu} , \, \lambda _{1} )
B_{2} ( p_{2}^{\mu} , \, \lambda _{2} ) \to B_{3} ( p_{3}^{\mu} , \,
\lambda _{3}) B_{4} ( p_{4}^{\mu} , \, \lambda _{4} )$, where the
momenta $p_{a}^{\mu}$ ($a= 1$, $2$, $3$, and $4$) satisfy $p_{1}^{\mu}
+ p_{2}^{\mu} = p_{3}^{\mu} + p_{4}^{\mu}$ and $\lambda _{a}$ is the
helicity of the baryon $B_{a}$.  Since we consider scatterings in the
center-of-mass frame, we can write the three-momenta as $\bm{p} \equiv
\bm{p}_{1} = - \bm{p}_{2}$ and $\bm{p}^{\prime} \equiv \bm{p}_{3} = -
\bm{p}_{4}$.  Without loss of generality, we can choose the
coordinates such that
\begin{equation}
  \bm{p} = ( 0 , \, 0 , \, p ) ,
  \quad 
  \bm{p}^{\prime} = ( p^{\prime} \sin \theta , \, 0 , \,
  p^{\prime} \cos \theta ) ,
\end{equation}
with the scattering angle $\theta$.  The mass of the baryon $B_{a}$
is expressed as $m_{a}$.

We calculate the partial-wave matrix elements of the interaction
$V_{\alpha}$ by following the Jacob--Wick
formulation~\cite{Jacob:1959at}, where $\alpha$ specifies the quantum
numbers of the system (see below).  First, according to Feynman
diagrams, we calculate the interactions in terms of the helicity
eigenstates as $V ( \bm{p}^{\prime} , \, \lambda _{3} , \, \lambda
_{4} , \, \bm{p} , \, \lambda _{1} , \, \lambda _{2} )$, whose
explicit forms are shown in the main part of this manuscript.  Then,
the interactions are projected to the total angular momentum $J$ as
\begin{align}
  & V^{J} ( p^{\prime} , \, \lambda _{3} , \, \lambda
  _{4} , \, p , \, \lambda _{1} , \, \lambda _{2} )
  \notag \\
  & = \frac{ \kappa ( p^{\prime} , \, p )}{2}
  \int _{-1}^{1} d \cos \theta \,
  d^{J}_{\lambda _{1} - \lambda _{2} \, \lambda _{3} - \lambda _{4}}
  ( \theta )
  \notag \\ & \phantom{=}
  \times V ( \bm{p}^{\prime} , \, \lambda _{3} , \,
  \lambda _{4} , \, \bm{p} , \, \lambda _{1} , \, \lambda _{2} ) ,
\end{align}
where $d_{m^{\prime} m}^{j}$ is the Wigner $d$-matrix and the factor
$\kappa ( p^{\prime} , \, p )$ is defined as
\begin{equation}
  \kappa ( p^{\prime} , \, p ) \equiv
  \sqrt{\frac{m_{1} m_{2} m_{3} m_{4}}
    {E_{1} ( p ) E_{2} ( p ) E_{3} ( p^{\prime} ) E_{4} ( p^{\prime} )}} ,
\end{equation}
with $E_{a} ( p ) \equiv \sqrt{p^{2} + m_{a}^{2}}$.  The factor
$\kappa$ was introduced so as to satisfy the optical theorem with the
correct coefficients.

Finally, the interaction used for the Lippmann--Schwinger
equation~\eqref{eq:LSeq_full} is obtained as
\begin{align}
  V_{\alpha} ( p^{\prime} , \, p )
  & = \sum _{\lambda _{1} , \lambda _{2} ,
    \lambda _{3} , \lambda _{4}}
  \frac{\sqrt{( 2 L + 1 )( 2 L^{\prime} + 1 )}}{2 J + 1}
  \notag \\
  & \phantom{=} \times
  \langle j_{3} \, j_{4} \, \lambda _{3} \,
  - \lambda _{4} | S^{\prime} \, S_{z}^{\prime} \rangle
  \langle L^{\prime} \, S^{\prime} \, 0 \, S_{z}^{\prime} | J \, S_{z}^{\prime} \rangle
  \notag \\
  & \phantom{=} \times
  \langle j_{1} \, j_{2} \, \lambda _{1} \, - \lambda _{2} | S S_{z} \rangle
  \langle L \, S \, 0 \, S_{z} | J \, S_{z} \rangle
  \notag \\
  & \phantom{=} \times
  V^{J} ( p^{\prime} , \, \lambda _{3} , \, \lambda _{4}
  , \, p , \, \lambda _{1} , \, \lambda _{2}) ,
\end{align}
where $j_{a}$ is the spin of the baryon $B_{a}$, $L^{( \prime )}$ and
$S^{( \prime )}$ are the orbital angular momentum and spin in the
initial (final) state, respectively, $S_{z} \equiv \lambda _{1} -
\lambda _{2}$, and $S_{z}^{\prime} \equiv \lambda _{3} - \lambda
_{4}$.

We note that the orbital angular momentum $L$ and spin $S$ may take
different values in the initial and final states as long as the total
angular momentum $J$ and parity\footnote{Note that all baryons in this
  study have positive parity.} $P = ( - 1 )^{L}$ of the system is
conserved.  An important example is the mixing of the $S$- and
$D$-wave components for the $N \Omega$ system.  In this sense, the
quantum number is specified as $\alpha = ( J , \, P , \, L^{\prime} ,
\, S^{\prime} , \, L , \, S )$.

\section{\boldmath Correlated two-meson exchange}
\label{app:corr}

In this Appendix we summarize our formulation of the correlated
two-meson exchange, for which we concentrate on the exchange of the
scalar-isoscalar channel $( J^{P} , \, I ) = ( 0^{+} , \, 0 )$.  The
contribution of the correlated two-meson exchange was expressed as
$\mathcal{V}_{\text{S}, \text{2M}}$ in Eq.~\eqref{eq:VB}.  According
to the dispersion relation~\eqref{eq:disp}, to calculate the
correlated two-meson exchange taking place in the region $t < 0$, we
may consider the same amplitude but in $t > 4 m_{\pi}^{2}$, which can
be achieved in the $N \bar{N} \to \Omega \bar{\Omega}$ reaction as
shown in Fig.~\ref{fig:NNbar}.

Let us formulate the $N \bar{N} \to \Omega \bar{\Omega}$ reaction in
Fig.~\ref{fig:NNbar}.  We fix the nucleon momenta $p_{N}^{\mu} = (
\sqrt{t} / 2 , \, \bm{p} )$ and $p_{\bar{N}}^{\mu} = ( \sqrt{t} / 2 ,
\, - \bm{p} )$ with $\bm{p} = ( 0, \, 0, \, p )$, and the $\Omega$
momenta $p_{\Omega}^{\mu} = ( \sqrt{t} / 2 , \, \bm{p}^{\prime} )$ and
$p_{\bar{\Omega}}^{\mu} = ( \sqrt{t} / 2 , \, - \bm{p}^{\prime} )$
with $\bm{p}^{\prime} = ( p^{\prime} \sin \theta , \, 0, \, p^{\prime}
\cos \theta)$ and the scattering angle $\theta$.  Because we
concentrate on the scalar channel, the scattering amplitude of the $N
\bar{N} \to \Omega \bar{\Omega}$ reaction can be evaluated as the
matrix element of the corresponding $T$-matrix $\hat{T}^{J = 0}$,
which contains $\mathcal{V}_{\text{S}, \text{2M}}$ according to
crossing symmetry:
\begin{align}
  & \langle \Omega \bar{\Omega}
  ( \bm{p}^{\prime} , \, \lambda _{\Omega} , \, \lambda _{\bar{\Omega}} ) |
  \hat{T}^{J = 0}
  | N \bar{N} ( \bm{p} , \, \lambda _{N} , \, \lambda _{\bar{N}} ) \rangle
  \notag \\
  & = \delta _{\lambda _{N} \, \lambda _{\bar{N}}}
  \delta _{\lambda _{\Omega} \, \lambda _{\bar{\Omega}}} \times 
  \bar{v}_{N} ( - \bm{p} , \, \lambda _{N} ) u_{N} ( \bm{p} , \, \lambda _{N} )
  \notag \\
  & \phantom{=} \times \left [ \mathcal{V}_{\rm S} ( t )
  \bar{u}_{\Omega}^{\mu} ( \bm{p}^{\prime} , \, \lambda _{\Omega} ) 
  v_{\Omega \, \mu} ( - \bm{p}^{\prime} , \, \lambda _{\Omega} ) 
  \phantom{\frac{q_{\mu} q_{\nu}}{m_{\Omega}^{2}}} \right .
  \notag \\
  & \phantom{= \times [} \left . + \mathcal{V}_{\rm 2M} ( t )
    \frac{P_{\mu} P_{\nu}}{m_{\Omega}^{2}}
  \bar{u}_{\Omega}^{\mu} ( \bm{p}^{\prime} , \, \lambda _{\Omega} ) 
  v_{\Omega}^{\nu} ( - \bm{p}^{\prime} , \, \lambda _{\Omega} ) \right ] ,
  \label{eq:Vcorr_TJ0}
\end{align}
where $P^{\mu} \equiv p_{N}^{\mu} + p_{\bar{N}}^{\mu} = ( \sqrt{t} ,
\, \bm{0})$ and the constraints $\lambda _{N} = \lambda _{\bar{N}}$
and $\lambda _{\Omega} = \lambda _{\bar{\Omega}}$ are necessary to
construct $J = 0$.  Here we note some relations for the spinors:
\begin{equation}
  \bar{v}_{N} ( - \bm{p} , \, \pm 1/2 ) 
  u_{N} ( \bm{p}, \, \pm 1/2 ) 
  = \frac{i p}{m_{N}} ,
  \label{eqA:vNuN}
\end{equation}
\begin{align}
  \bar{u}_{\Omega}^{\mu} ( \bm{p}^{\prime} , \, \pm 3/2 ) 
  v_{\bar{\Omega} \, \mu} ( - \bm{p}^{\prime} , \, \pm 3/2 ) 
  = \frac{i p^{\prime}}{m_{\Omega}} ,
  \label{eqA:uOvOA}
\end{align}
\begin{align}
  \bar{u}_{\Omega}^{\mu} ( \bm{p}^{\prime} , \, \pm 1/2 ) 
  v_{\bar{\Omega} \, \mu} ( - \bm{p}^{\prime} , \, \pm 1/2 )
  = - \frac{i p^{\prime}}{3 m_{\Omega}}
  \left ( 1 + \frac{4 p^{\prime \, 2}}{m_{\Omega}^{2}} \right ) ,
  \label{eqA:uOvOB}
\end{align}
\begin{align}
  \frac{P_{\mu} P_{\nu}}{m_{\Omega}^{2}}
  \bar{u}_{\Omega}^{\mu} ( \bm{p}^{\prime} , \, \pm 3/2 ) 
  v_{\Omega}^{\nu} ( - \bm{p}^{\prime} , \, \pm 3/2 ) 
  = 0 ,
  \label{eqA:uOvOC}
\end{align}
\begin{align}
  \frac{P_{\mu} P_{\nu}}{m_{\Omega}^{2}}
  \bar{u}_{\Omega}^{\mu} ( \bm{p}^{\prime} , \, \pm 1/2 ) 
  v_{\Omega}^{\nu} ( - \bm{p}^{\prime} , \, \pm 1/2 ) 
  = - \frac{2 i p^{\prime \, 3} t}{3 m_{\Omega}^{5}} ,
  \label{eqA:uOvOD}
\end{align}
where double-sign corresponds.

To calculate $\mathcal{V}_{\text{S}, \text{2M}}$ in the region $t < 0$
via the dispersion relation~\eqref{eq:disp}, we need $\text{Im}
\mathcal{V}_{\text{S}, \text{2M}} ( t )$ in $t > 4 m_{\pi}^{2}$.  For
this purpose, we first recall the unitarity of the $S$-matrix:
$\hat{S}^{\dagger} \hat{S} = 1$.  Expressing this relation in terms
of the $T$-matrix of the $N \bar{N} \to \Omega \bar{\Omega}$ reaction
in the scalar channel, we have
\begin{align}
  & i \langle \Omega \bar{\Omega} | \hat{T}^{J = 0} | N \bar{N} \rangle
  - i \langle \Omega \bar{\Omega} | \hat{T}^{J = 0 \, \dagger} | N \bar{N} \rangle
  \notag \\
  & = \sum _{n} \rho _{n} ( t )
  \theta ( t - m_{\text{th} ( n )}^{2} )
  \langle \Omega \bar{\Omega} | \hat{T}^{J = 0 \, \dagger} | n \rangle
  \langle n | \hat{T}^{J = 0} | N \bar{N} \rangle ,
  \label{eqA:TTdagger}
\end{align}
where we omitted the parameters ($\bm{p}^{\prime}$, $\lambda
_{\Omega}$, $\lambda _{\bar{\Omega}}$) for the $\Omega \bar{\Omega}$
state and ($\bm{p}$, $\lambda _{N}$, $\lambda _{\bar{N}}$) for the $N
\bar{N}$ state.  In the right-hand side, $n = \pi \pi$, $K \bar{K}$,
$\eta \eta$, $\ldots$ denotes possible physical channels,
$m_{\text{th} ( n )}$ is its threshold, and $\rho _{n} ( t )$ is its
phase space.  In particular,
\begin{equation}
  \rho _{P \bar{P}} ( t ) = \frac{N_{P \bar{P}}}{8 \pi}
  \sqrt{\frac{t - 4 m_{P}^{2}}{t}} ,
\end{equation}
for $n = P \bar{P} = \pi \pi , \, K \bar{K} , \, \eta \eta$, with the
mass of the pseudoscalar meson $m_{P}$ and the symmetry factor for
identical particles: $N_{\pi \pi} = N_{\eta \eta} = 1/2$ and $N_{K
  \bar{K}} = 1$.  Therefore, by using the relation in
Eq.~\eqref{eq:Vcorr_TJ0} and $\langle A | \hat{T}^{\dagger} | B
\rangle = \langle B | \hat{T} | A \rangle ^{\ast}$, we can rewrite
\eqref{eqA:TTdagger} as
\begin{align}
  & \delta _{\lambda _{N} \, \lambda _{\bar{N}}}
  \delta _{\lambda _{\Omega} \, \lambda _{\bar{\Omega}}} \times 
  \bar{v}_{N} ( - \bm{p} , \, \lambda _{N} ) u_{N} ( \bm{p} , \, \lambda _{N} )
  \notag \\
  & \times \left [ - 2 \, \text{Im} \mathcal{V}_{\rm S} ( t )
  \bar{u}_{\Omega}^{\mu} ( \bm{p}^{\prime} , \, \lambda _{\Omega} ) 
  v_{\Omega \, \mu} ( - \bm{p}^{\prime} , \, \lambda _{\Omega} ) 
  \phantom{\frac{q_{\mu} q_{\nu}}{m_{\Omega}^{2}}} \right .
  \notag \\
  & \phantom{\times [} \left . - 2 \, \text{Im} \mathcal{V}_{\rm 2M} ( t )
    \frac{P_{\mu} P_{\nu}}{m_{\Omega}^{2}}
  \bar{u}_{\Omega}^{\mu} ( \bm{p}^{\prime} , \, \lambda _{\Omega} ) 
  v_{\Omega}^{\nu} ( - \bm{p}^{\prime} , \, \lambda _{\Omega} ) \right ] 
\notag \\
  & = \sum _{n} \rho _{n} ( t )
  \theta ( t - m_{\text{th} ( n )}^{2} )
  \langle n | \hat{T}^{J = 0} | \Omega \bar{\Omega} \rangle ^{\ast}
  \langle n | \hat{T}^{J = 0} | N \bar{N} \rangle .
  \label{eq:ImV_gen}
\end{align}
This is a general formula to calculate $\text{Im}
\mathcal{V}_{\text{S}, \text{2M}} ( t )$ in the region $t > 4
m_{\pi}^{2}$.  In the equation the magnitudes of the momenta of $N$
and $\Omega$, $p ( t )$ and $p^{\prime} ( t )$, respectively, take
their on-shell value
\begin{equation}
  p ( t ) = \sqrt{\frac{t}{4} - m_{N}^{2}} ,
  \quad 
  p^{\prime} ( t ) = \sqrt{\frac{t}{4} - m_{\Omega}^{2}} .
  \label{eq:ponshell}
\end{equation}
Note that the unitarity relation~\eqref{eq:ImV_gen} is defined above
the kinematic threshold of the $N \bar{N} \to \Omega \bar{\Omega}$
reaction, i.e., $t > 4 m_{\Omega}^{2}$.  However, one can perform the
analytic continuation into the pseudophysical region $t < 4
m_{\Omega}^{2}$, where $p^{\prime} ( t )$ [and $p ( t )$ in $t < 4
  m_{N}^{2}$] is pure imaginary.

In general, $n$ in Eq.~\eqref{eq:ImV_gen} should run all possible
physical channels, but it is well known that the scalar-isoscalar
channel in the region $4 m_{\pi}^{2} < t \lesssim 1 \gev ^{2}$ is
dominated by the contributions from the dynamics of $\pi \pi$ and
coupled channels of two pseudoscalar mesons.  Therefore, below we
restrict the summation in Eq.~\eqref{eq:ImV_gen} to the physical $\pi
\pi$, $K \bar{K}$, and $\eta \eta$ states.

Now our task is to calculate the scattering amplitudes of the $N
\bar{N} \to P \bar{P}$ and $\Omega \bar{\Omega} \to P \bar{P}$ ($P
\bar{P} = \pi \pi$, $K \bar{K}$, $\eta \eta$) reactions in the
scalar-isoscalar channel.  Here, to simplify the evaluation of
  the $N \Omega$ interaction from the $N \bar{N} \to \Omega
  \bar{\Omega}$ amplitude, $N \bar{N}$ is in the particle basis,
i.e., $N \bar{N} = p \bar{p}$ or $n \bar{n}$, while the $P \bar{P}$ is
in the isospin basis with $I = 0$:
\begin{align}
  | \pi \pi ( \bm{k} ) \rangle = - \frac{1}{\sqrt{3}}
  | & \pi ^{+} ( \bm{k} ) \pi ^{-} ( - \bm{k} )
  + \pi ^{0} ( \bm{k} ) \pi ^{0} ( - \bm{k} )
  \notag \\
  & 
  + \pi ^{-} ( \bm{k} ) \pi ^{+} ( - \bm{k} )\rangle , 
\end{align}
\begin{align}
  | K \bar{K} ( \bm{k} ) \rangle = - \frac{1}{\sqrt{2}}
  | K^{+} ( \bm{k} ) K^{-} ( - \bm{k} )
  + K^{0} ( \bm{k} ) \bar{K}^{0} ( - \bm{k} ) \rangle , 
\end{align}
\begin{align}
  | \eta \eta ( \bm{k} ) \rangle = 
  | \eta ( \bm{k} ) \eta ( - \bm{k} ) \rangle ,
\end{align}
where $\bm{k}$ is the relative momentum of mesons.  We explicitly
write the off-shell amplitudes of the $N \bar{N} \to P \bar{P}$
and $\Omega \bar{\Omega} \to P \bar{P}$ reactions as
\begin{align}
  T_{N \bar{N} \to P \bar{P}} ( t , \, k , \, p , \, \lambda _{N})
  & = \langle P \bar{P} ( \bm{k} ) | \hat{T}^{J = 0} | N \bar{N}
  ( \bm{p} , \, \lambda _{N} , \, \lambda _{N} ) \rangle ,
\end{align}
\begin{align}
  T_{\Omega \bar{\Omega} \to P \bar{P}}
  ( t , \, k , \, p^{\prime} , \, \lambda _{\Omega} )
  = \langle P \bar{P} ( \bm{k} ) | \hat{T}^{J = 0} | \Omega \bar{\Omega}
  ( \bm{p}^{\prime} , \, \lambda _{\Omega} , \, \lambda _{\Omega} ) \rangle .
\end{align}
Helicities are constrained as $\lambda _{N} = \lambda _{\bar{N}}$ and
$\lambda _{\Omega} = \lambda _{\bar{\Omega}}$ so as to construct $J =
0$. Because of this $S$-wave nature, the left-hand-side depends only
on the magnitude of the momenta.  Owing to the parity invariance of
the underlying strong interaction, the amplitudes $T_{N \bar{N} \to P
  \bar{P}}$ and $T_{\Omega \bar{\Omega} \to P \bar{P}}$ have relations
\begin{equation}
  T_{N \bar{N} \to P \bar{P}} ( t , \, k , \, p , \, \lambda _{N} )
  = T_{N \bar{N} \to P \bar{P}} ( t , \, k , \, p , \, - \lambda _{N} ) ,
\end{equation}
\begin{equation}
  T_{\Omega \bar{\Omega} \to P \bar{P}}
  ( t , \, k , \, p^{\prime} , \, \lambda _{\Omega} )
  = T_{\Omega \bar{\Omega} \to P \bar{P}}
  ( t , \, k , \, p^{\prime} , \, - \lambda _{\Omega} ) .
\end{equation}
Therefore, while $T_{N \bar{N} \to P \bar{P}}$ does not depend on
$\lambda _{N}$, $T_{\Omega \bar{\Omega} \to P \bar{P}}$ has two
independent components of $\lambda _{\Omega} = 3/2$ and $1/2$.  Note
that momenta $p$, $p^{\prime}$, and $k$ are independent of $t$ in the
off-shell amplitudes.  One can easily obtain the on-shell amplitudes
of the $N \bar{N} \to P \bar{P}$ and $\Omega \bar{\Omega} \to P
\bar{P}$ reactions by putting on-shell momenta of the baryons $q ( t
)$ and $q^{\prime} ( t )$ in Eq.~\eqref{eq:ponshell} and
\begin{align}
  k ( t ) = \sqrt{\frac{t}{4} - m_{P}^{2}} ,
\end{align}
respectively.

\begin{figure}[!t]
  \centering
  \PsfigII{0.215}{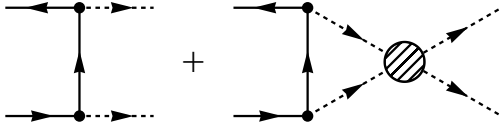}
  \caption{Diagrammatic equation for the $N \bar{N} \to P \bar{P}$
    and $\Omega \bar{\Omega} \to P \bar{P}$ scattering amplitudes.
    Solid and dashed lines represent baryons and mesons, respectively.
    Shaded circle denotes the correlation of two mesons.}
  \label{fig:NNT}
\end{figure}

The $N \bar{N}$, $\Omega \bar{\Omega} \to P \bar{P}$ amplitudes
are calculated according to the diagram in Fig.~\ref{fig:NNT}.  To
this end, we label the two meson channels $\pi \pi$, $K \bar{K}$, and
$\eta \eta$ as $j = 1$, $2$, and $3$, respectively, and we project the
Born term, i.e., the first term in Fig.~\ref{fig:NNT}, into the scalar
channel as
\begin{align}
  & V_{j}^{(N)} ( t , \, k , \, p )
  = \frac{1}{2} \int _{-1}^{1} d \cos \theta _{k} \langle j ( \bm{k} ) | \hat{V}
  | N \bar{N} ( \bm{p} , \, \lambda _{N} , \, \lambda _{N} ) \rangle ,
\end{align}
\begin{align}
  & V_{j}^{( \Omega )} ( t , \, k , \, p^{\prime} , \, \lambda _{\Omega} )
  \notag \\ &
  = \frac{1}{2} \int _{-1}^{1} d \cos \theta _{k}^{\prime}
  \langle j ( \bm{k} ) | \hat{V}
  | \Omega \bar{\Omega} ( \bm{p}^{\prime} , \, \lambda _{\Omega} , \,
  \lambda _{\Omega} ) \rangle ,
\end{align}
where $\theta _{k}^{( \prime )}$ is the angle between $\bm{p}^{(
  \prime )}$ and $\bm{k}$.  The explicit forms of the matrix elements
of $\hat{V}$ are shown in Appendix~\ref{app:explicit}.  Then, we
evaluate the diagram in Fig.~\ref{fig:NNT} according to the
prescription by Blankenbecler--Sugar~\cite{Blankenbecler:1965gx} as
\begin{align}
  & T_{N \bar{N} \to j} ( t , \, k , \, p )
  = V_{j}^{(N)} ( t , \, k , \, p )
  \notag \\
  & \quad + \sum _{l=1}^{3} N_{l} \int _{0}^{\infty} d k^{\prime}
  \frac{k^{\prime \, 2}}{2 \pi ^{2}}
  \frac{T_{j l}^{\rm (2m)} ( t ) 
    V_{l}^{(N)} ( t , \, k^{\prime} , \, p )}{\omega _{P ( l )} ( k^{\prime} )
    [ t - 4 \omega _{P ( l )} ( k^{\prime} )^{2} ]} ,
  \label{eq:TNNbar_final}
\end{align}
\begin{align}
  & T_{\Omega \bar{\Omega} \to j} ( t , \, k , \, p^{\prime} , \, \lambda _{\Omega} )
  = V_{j}^{( \Omega )} ( t , \, k , \, p^{\prime} , \, \lambda _{\Omega} )
  \notag \\
  & \quad + \sum _{l=1}^{3} N_{l} \int _{0}^{\infty} d k^{\prime}
  \frac{k^{\prime \, 2}}{2 \pi ^{2}}
  \frac{T_{j l}^{\rm (2m)} ( t ) 
    V_{l}^{( \Omega )} ( t , \, k^{\prime} , \, p^{\prime} , \, \lambda _{\Omega})}
       {\omega _{P ( l )} ( k^{\prime} )
    [ t - 4 \omega _{P ( l )} ( k^{\prime} )^{2} ]} ,
  \label{eq:TOObar_final}
\end{align}
where $\omega _{P ( l )} ( k ) \equiv \sqrt{k^{2} + m_{P ( l )}^{2}}$
with $P ( l )$ being the meson in $l$th channel and $T_{j l}^{\rm
    (2m)}$ is the scalar-isoscalar $l \to j$ meson--meson scattering
  amplitude.  In general, $T_{j l}^{\rm (2m)}$ should be a off-shell
  amplitude and hence depends on the relative momenta $k$ and
  $k^{\prime}$ as well.  In the present study, we employ the so-called
  chiral unitary approach to describe $T_{j l}^{\rm (2m)}$ together
  with the on-shell approximation as explained in
  Appendix~\ref{app:meson}, so $T_{j l}^{\rm (2m)}$ is a function only
  of $t$.

\begin{figure}[!t]
  \centering
  \Psfig{8.6cm}{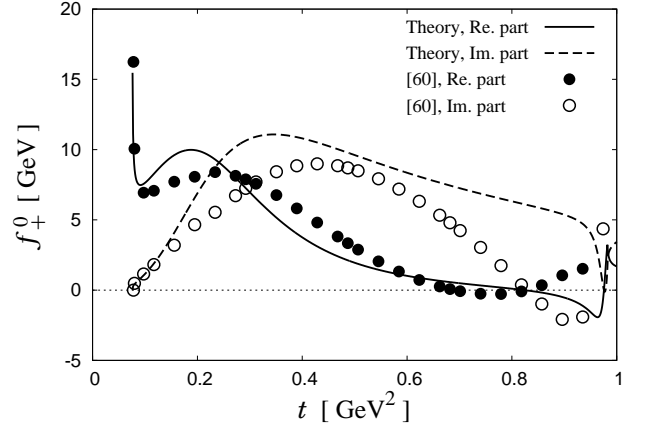}
  \caption{Frazer--Fulco amplitude $f_{+}^{0}$ for the $N \bar{N} \to
    \pi \pi$ reaction.  Points represent quasiempirical values taken
    from Ref.~\cite{Dumbrajs:1983jd}.}
  \label{fig:FF}
\end{figure}

With Appendices~\ref{app:explicit} and \ref{app:meson}, all the
ingredients in the above amplitudes are determined.  We can check how
the $N \bar{N} \to \pi \pi$ amplitude in the present formulation works
by calculating the Frazer--Fulco amplitude for the $N \bar{N} \to \pi
\pi$ reaction~\cite{Frazer:1960zza}
\begin{equation}
  f_{+}^{0} ( t ) = \frac{i p ( t ) m_{N}}{4 \sqrt{3} \pi} 
  T_{N \bar{N} \to \pi \pi} ( t , \, k ( t ) , \, p ( t ) ) ,
\end{equation}
where the factor is due to the transition to the Frazer-Fulco
  amplitude in isospin basis.  The result is shown in
Fig.~\ref{fig:FF} together with quasiempirical values taken from
Ref.~\cite{Dumbrajs:1983jd}. The comparison indicates that our
approach reproduces the quasiempirical values semi-quantitatively
well.

\begin{figure}[!t]
  \centering
  \PsfigII{0.215}{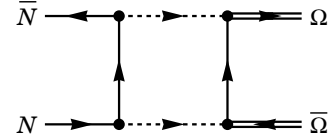}
  \caption{Feynman box diagram for the $N \bar{N} \to \Omega
    \bar{\Omega}$ reaction.  Solid and double lines in
    the intermediate state represent baryons, and dashed lines
    mesons.}
  \label{fig:NNbox}
\end{figure}

Finally, we evaluate $\text{Im} \mathcal{V}_{\text{S}, \text{2M}} ( t
)$ in $t > 4 m_{\pi}^{2}$ from the on-shell amplitudes in
Eqs.~\eqref{eq:TNNbar_final} and \eqref{eq:TOObar_final} and the
relation in Eq.~\eqref{eq:ImV_gen}.  We express the right-hand side of
Eq.~\eqref{eq:ImV_gen}, i.e., the sum of the products of the phase
space $\rho _{j}$ and amplitudes of the $j \to \Omega \bar{\Omega}$
and $N \bar{N} \to j$ reactions, as $F ( t , \, \lambda _{\Omega} )$.
We note that the product $T_{\Omega \bar{\Omega} \to j} ^{\ast} T_{N
  \bar{N} \to j}$ contains the uncorrelated contributions depicted as
the box diagram in Fig.~\ref{fig:NNbox} which eventually causes the
double counting in the $N\Omega$ interaction with the $\eta$ exchange
term and with the box contributions of the inelastic channel in
Section~\ref{sec:2H}.  We must cancel this double counting by
subtracting the product of the Born terms.  As a result, $F ( t , \,
\lambda _{\Omega} )$ is
\begin{align}
  & F ( t , \, \lambda _{\Omega} ) \equiv
  \sum _{j = 1}^{3} \rho _{j} ( t ) \theta ( t - m_{\text{th} ( j )}^{2} )
  \notag \\
  & \quad \times
  \Big [ T_{\Omega \bar{\Omega} \to j} ( t , \, k ( t ) , \, p^{\prime} ( t ) , \,
  \lambda _{\Omega} )^{\ast}
  T_{N \bar{N} \to j} ( t , \, k ( t ) , \, p ( t ) ) 
  \notag \\
  & \quad \quad 
    - V_{j}^{( \Omega )} ( t , \, k ( t ) , \, p^{\prime} ( t ) , \,
  \lambda _{\Omega} )^{\ast} V_{j}^{(N)} ( t , \, k ( t ) , \, p ( t ) ) 
\Big ] .
\end{align}
With this and the formulae in Eqs.~\eqref{eqA:vNuN},
\eqref{eqA:uOvOA}, \eqref{eqA:uOvOB}, \eqref{eqA:uOvOC}, and
\eqref{eqA:uOvOD}, we have
\begin{align}
  & \text{Im} \mathcal{V}_{\rm S} ( t )
  \notag \\ & 
  = - \frac{F ( t , \, 3/2 )}{2 \bar{u}_{\Omega}^{\mu} ( \bm{p}^{\prime} , \, 3/2 )
    v_{\bar{\Omega} \, \mu} ( - \bm{p}^{\prime} , \, 3/2 )
    \bar{v}_{N} ( - \bm{p} , \, 1/2 )
    u_{N} ( \bm{p}, \, 1/2 )}
  \notag \\ & 
  = - \frac{2 m_{N} m_{\Omega} F ( t , \, 3/2 )}{\sqrt{( 4 m_{N}^{2} - t )
      ( 4 m_{\Omega}^{2} - t )}} ,
\end{align}
\begin{align}
  \text{Im} \mathcal{V}_{\rm 2M} ( t ) 
  = & \frac{m_{\Omega}^{2}}{P_{\mu} P_{\nu}
    \bar{u}_{\Omega}^{\mu} ( \bm{p}^{\prime} , \, 1/2 )
    v_{\Omega}^{\nu} ( - \bm{p}^{\prime} , \, 1/2 )}
  \notag \\ &
  \times
  \left [ - \frac{F ( t, \, 1/2 )}{2 \bar{v}_{N} ( - \bm{p} , \, 1/2 )
      u_{N} ( \bm{p}, \, 1/2 )} \right .
    \notag \\ &
  \left . \phantom{\frac{F}{\bar{v}_{N}}}
  - \text{Im} \mathcal{V}_{\text{S}} ( t ) 
  \bar{u}_{\Omega}^{\mu} ( \bm{p}^{\prime} , \, 1/2 )
  v_{\Omega \, \mu} ( - \bm{p}^{\prime} , \, 1/2 ) \right ]
  \notag \\ = & 
- \frac{4 m_{N} m_{\Omega}^{3}}{\sqrt{( 4 m_{N}^{2} - t ) ( 4 m_{\Omega}^{2} - t )}
      ( 4 m_{\Omega}^{2} - t ) t}
\notag \\ & \times
    \left [ 3 m_{\Omega}^{2} F ( t , \, 1/2 ) 
      + ( t - 3 m_{\Omega}^{2} ) F ( t, \, 3/2 ) \right ] .
\end{align}

\begin{figure}[!t]
  \centering
  \Psfig{8.6cm}{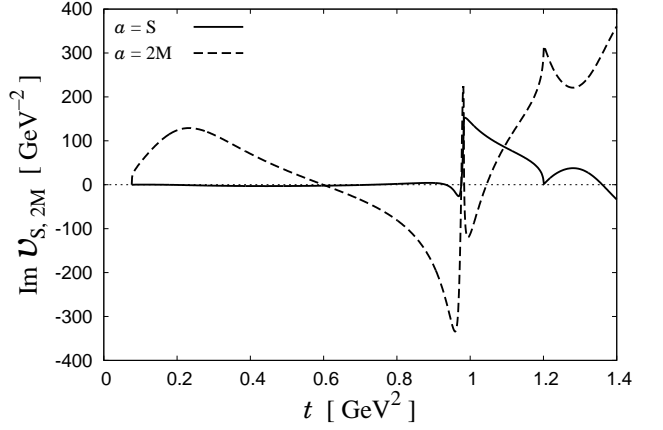}
  \caption{$\text{Im} \mathcal{V}_{\rm S}$ and $\text{Im}
    \mathcal{V}_{\rm 2M}$ in our model.}
  \label{fig:ImVcorr}
\end{figure}

In Fig.~\ref{fig:ImVcorr} we plot $\text{Im} \mathcal{V}_{\rm S}$ and
$\text{Im} \mathcal{V}_{\rm 2M}$ in our model as functions of $t > 4
m_{\pi}^{2}$.  As one can see, $\text{Im} \mathcal{V}_{\rm S}$ takes a
nonnegligible value only for $t \gtrsim 1 \gev ^{2}$.  This may be
interpreted as the exchanges of the $f_{0} (980)$ and correlated $K
\bar{K}$ states.  On the other hand, $\text{Im} \mathcal{V}_{\rm 2M}$
has a contribution at just above the threshold $t = 4 m_{\pi}^{2}$ as
well, reflecting the contributions from the broad ``$\sigma$'' meson
and correlated $\pi \pi$.

Then, we utilize the dispersion relation~\eqref{eq:disp} to evaluate
$\mathcal{V}_{\text{S}, \text{2M}} ( t )$ in $t < 0$, where the
correlated two-meson exchange takes place in the $N \Omega$ elastic
scattering.  Here we perform the integration with a cutoff $t_{c}$
instead of infinity, which we take $t_{c} = ( 1.2 \gev )^{2}$, which
is the upper boundary of the fit range of our $\pi \pi$-$K
\bar{K}$-$\eta \eta$ scattering amplitude in the chiral unitary
approach to the experimental $\pi \pi ( J = 0 , \, I = 0 )$ phase
shift (Appendix~\ref{app:meson}).

\section{\boldmath Explicit forms of the Born terms for the
  $N \bar{N}$, $\Omega \bar{\Omega} \to$ meson--meson reactions}
\label{app:explicit}

Here we show the explicit forms of the Born terms for the $N ( \bm{p}
, \, \lambda _{N} ) \bar{N} ( - \bm{p} , \, \lambda _{N} ) \to P (
\bm{k} ) \bar{P} ( - \bm{k} )$ and $\Omega ( \bm{p}^{\prime} , \,
\lambda _{\Omega} ) \bar{\Omega} ( - \bm{p}^{\prime} , \, \lambda
_{\Omega} ) \to P ( \bm{k} ) \bar{P} ( - \bm{k} )$ reactions,
where $P \bar{P} = \pi \pi$, $K \bar{K}$, and $\eta \eta$.  Here
$\bm{p}$, $\bm{p}^{\prime}$, and $\bm{k}$ are relative momenta of $N
\bar{N}$, $\Omega \bar{\Omega}$, and $P \bar{P}$, respectively,
for which we take, without loss of generality, $\bm{p} = ( 0 , \, 0 ,
\, p )$, $\bm{p}^{\prime} = ( 0 , \, 0 , \, p^{\prime} )$, and $\bm{k}
= ( k \sin \theta , \, 0 , \, k \cos \theta )$ throughout this
section. We define four-momenta $p_{N}^{\mu} = ( \sqrt{t} / 2 , \, \bm{p})$,
$p_{\bar{N}}^{\mu} = ( \sqrt{t} / 2 , \, - \bm{p})$, 
$p_{\Omega}^{\mu} = ( \sqrt{t} / 2 , \, \bm{p}^{\prime})$, 
$p_{\bar{\Omega}}^{\mu} = ( \sqrt{t} / 2 , \, - \bm{p}^{\prime})$,
$k_{P}^{\mu} \equiv k^{\mu} = ( \sqrt{t} / 2 , \, \bm{k})$, and
$k_{\bar{P}}^{\mu} \equiv k^{\prime \, \mu} = ( \sqrt{t} / 2 , \, - \bm{k})$.
Helicities of antiparticles are constrained as $\lambda
_{\bar{N}} = \lambda _{N}$ and $\lambda _{\bar{\Omega}} = \lambda
_{\Omega}$ so as to construct $J = 0$.

\subsection{\boldmath $N \bar{N} \to$ meson--meson}

First, the $N \bar{N} \to \pi \pi$, $K \bar{K}$, and $\eta \eta$ Born
terms are calculated as
\begin{align}
  & \langle \pi \pi ( \bm{k} ) | \hat{V} | N \bar{N} ( \bm{p} , \,
  \lambda _{N} , \, \lambda _{N} ) \rangle 
  \notag \\ &
  = \mathcal{F}_{\pi N N} \mathcal{V}_{N N N}
  + \mathcal{F}_{\pi N \Delta} \mathcal{V}_{N \Delta N}
  + ( \bm{k} \leftrightarrow - \bm{k} ) ,
\end{align}
\begin{align}
  & \langle K \bar{K} ( \bm{k} ) | \hat{V} | N \bar{N} ( \bm{p} , \,
  \lambda _{N} , \, \lambda _{N} ) \rangle 
  \notag \\ &
  = \mathcal{F}_{K N \Lambda} \mathcal{V}_{N \Lambda N}
  + \mathcal{F}_{K N \Sigma} \mathcal{V}_{N \Sigma N} 
  + \mathcal{F}_{K N \Sigma ^{\ast}} \mathcal{V}_{N \Sigma ^{\ast} N} ,
\end{align}
\begin{align}
  & \langle \eta \eta ( \bm{k} ) | \hat{V} | N \bar{N} ( \bm{p} , \,
  \lambda _{N} , \, \lambda _{N} ) \rangle 
  = \mathcal{F}_{\eta N N} \mathcal{V}_{N N N}
  + ( \bm{k} \leftrightarrow - \bm{k} ) .
\end{align}
The notation $( \bm{k}\leftrightarrow - \bm{k} )$ means to add the symmetrized contributions for the identical two-meson systems.
Here $\mathcal{F}_{P B B}$ and $\mathcal{F}_{P B D}$ are coupling
constants which are given by
\begin{equation}
  \mathcal{F}_{\pi N N}
  = - \frac{\sqrt{3} ( D + F )^{2}}{4 f_{\pi}^{2}} ,
  \quad
  \mathcal{F}_{\pi N \Delta}
  = - \frac{2}{\sqrt{3}} \frac{f_{P B D}^{2}}{m_{\pi}^{2}} ,
\end{equation}
\begin{equation}
  \mathcal{F}_{K N \Lambda}
  = - \frac{( D + 3 F )^{2}}{12 \sqrt{2} f_{K}^{2}} ,
  \quad
  \mathcal{F}_{K N \Sigma}
  = - \frac{3 ( D - F )^{2}}{4 \sqrt{2} f_{K}^{2}} ,
\end{equation}
\begin{equation}
  \mathcal{F}_{K N \Sigma ^{\ast}}
  = - \frac{1}{2 \sqrt{2}} \frac{f_{P B D}^{2}}{m_{\pi}^{2}} ,
  \quad
  \mathcal{F}_{\eta N N}
  = \frac{( D - 3 F )^{2}}{12 f_{\eta}^{2}} .
\end{equation}
Terms $\mathcal{V}_{N B N}$ and $\mathcal{V}_{N D N}$ are the
amplitudes of the octet- and decuplet-baryon exchange for the $N
\bar{N}$ scattering, respectively, as functions of $t$, $\bm{k}$,
and $\bm{p}$: 
\begin{align}
  & \mathcal{V}_{N B N} ( t , \, \bm{k} , \, \bm{p} )
  = - F ( k )^{2} 
  \bar{v}_{N} ( - \bm{p} , \, + 1/2 )
  \notag \\
  & \quad
  \times
  \Slash{k}^{\prime} \gamma _{5}
  S_{B} ( p_{N} - k )
  \Slash{k} \gamma _{5}
  u_{N} ( \bm{p} , \, + 1/2 )
  ,
\end{align}
where $S_{B} ( p )$ is the propagator of the octet baryon $B$
\begin{align}
  S_{B} ( p ) \equiv \frac{\Slash{p} + m_{B}}{ ( p^{\mu} )^{2} - m_{B}^{2}} ,
\end{align}
with its mass $m_{B}$, and
\begin{align}
  & \mathcal{V}_{N D N} ( t , \, \bm{k} , \, \bm{p} , \, \lambda _{N} )
  = F ( k )^{2} 
  \bar{v}_{N} ( - \bm{p} , \, + 1/2 )
  \notag \\
  & \quad
  \times
  k_{\mu}^{\prime}
  S_{D}^{\mu \nu} ( p_{N} - k )
  k_{\nu}
  u_{N} ( \bm{p} , \, + 1/2 )
  ,
\end{align}
where $S_{D} ( p )$ is the propagator of the decuplet baryon $D$
\begin{align}
  S_{D}^{\mu \nu} ( p )
  = & \frac{\Slash{p} + m_{D}}{( p^{\mu} )^{2} - m_{D}^{2}}
  \left [ g^{\mu \nu} - \frac{1}{3} \gamma ^{\mu} \gamma ^{\nu}
    - \frac{2 p^{\mu} p^{\nu}}{3 m_{D}^{2}} \right .
    \notag \\
    & \quad
    \left . + \frac{p^{\mu} \gamma ^{\nu} 
    - p^{\nu} \gamma ^{\mu}}{3 m_{D}} \right ]
    ,
\end{align}
with its mass $m_{D}$.  As for the form factor $F ( k )$ in the
  $N \bar{N}$, $\Omega \Omega \to P \bar{P}$ amplitudes, we employ the
  monopole type in Eq.~\eqref{eq:FF} and use the same value of the
  cutoff $\Lambda = 1 \gev$.  We do not include the width of the
decuplet baryons in the propagator.   Note that both $\mathcal{V}_{N B
  N}$ and $\mathcal{V}_{N D N}$ do not depend on the helicity $\lambda
_{N}$, so we take $\lambda _{N} = +1/2$ here.

\subsection{\boldmath $\Omega \bar{\Omega} \to$ meson--meson}

Next, the $\Omega \bar{\Omega} \to \pi \pi$, $K \bar{K}$, and $\eta
\eta$ Born terms are calculated as
\begin{align}
  & \langle \pi \pi ( \bm{k} ) | \hat{V} | \Omega \bar{\Omega} 
  ( \bm{p}^{\prime} , \, \lambda _{\Omega} , \, \lambda _{\Omega} ) \rangle 
  = 0 ,
  \label{eq:pipiOmegaOmega}
\end{align}
\begin{align}
  & \langle K \bar{K} ( \bm{k} ) | \hat{V} | \Omega \bar{\Omega} 
  ( \bm{p}^{\prime} , \, \lambda _{\Omega} , \, \lambda _{\Omega} ) \rangle 
  \notag \\ &
  = \mathcal{F}_{K \Xi \Omega} \mathcal{V}_{\Omega \Xi \Omega}
  + \mathcal{F}_{K \Xi ^{\ast} \Omega} \mathcal{V}_{\Omega \Xi ^{\ast} \Omega} ,
\end{align}
\begin{align}
  & \langle \eta \eta ( \bm{k} ) | \hat{V} | \Omega \bar{\Omega} 
  ( \bm{p}^{\prime} , \, \lambda _{\Omega} , \, \lambda _{\Omega} ) \rangle 
  = \mathcal{F}_{\eta \Omega \Omega} \mathcal{V}_{\Omega \Omega \Omega}
  + ( \bm{k} \leftrightarrow - \bm{k} ).
\end{align}
Here $\mathcal{F}_{P B B}$ and $\mathcal{F}_{P B D}$ are coupling
constants defined as
\begin{equation}
  \mathcal{F}_{K \Xi \Omega}
  = \sqrt{2} \frac{f_{P B D}^{2}}{m_{\pi}^{2}} ,
  \quad
  \mathcal{F}_{K \Xi ^{\ast} \Omega}
  = - \frac{\sqrt{2} f_{P D D}^{2}}{3 m_{\pi}^{2}} ,
\end{equation}
\begin{equation}
  \mathcal{F}_{\eta \Omega \Omega}
  = \frac{2 f_{P D D}}{3 m_{\pi}^{2}} ,
\end{equation}
Terms $\mathcal{V}_{\Omega B \Omega}$ and $\mathcal{V}_{\Omega D \Omega}$ are the
amplitudes of the octet- and decuplet-baryon exchange for the $\Omega
\bar{\Omega}$ scattering, respectively, as functions of $t$, $\bm{k}$,
$\bm{p}^{\prime}$, and $\lambda _{N}$:
\begin{align}
  & \mathcal{V}_{\Omega B \Omega}
  ( t , \, \bm{k} , \, \bm{p}^{\prime} , \, \lambda _{\Omega} )
  =
  F ( k )^{2} \bar{v}_{\Omega}^{\mu} ( - \bm{p}^{\prime} , \, \lambda _{\Omega} )
  \notag \\
  & \quad
  \times k^{\prime}_{\mu} S_{B} ( p_{\Omega} - k ) k_{\nu} 
  u_{\Omega}^{\nu} ( \bm{p}^{\prime} , \, \lambda _{\Omega} ) ,
\end{align}
\begin{align}
  & \mathcal{V}_{\Omega D \Omega}
  ( t , \, \bm{k} , \, \bm{p}^{\prime} , \, \lambda _{\Omega} )
  = F ( k )^{2}
  \bar{v}_{\Omega \, \mu} ( - \bm{p}^{\prime} , \, \lambda _{\Omega} )
  \notag \\
  & \quad
  \times \Slash{k}^{\prime} \gamma _{5}
  S_{D}^{\mu \nu} ( p_{\Omega} - k ) 
  \Slash{k} \gamma _{5}
  u_{\Omega \, \nu} ( \bm{p}^{\prime} , \, \lambda _{\Omega} ) ,  
\end{align}
Note that the amplitudes depend on the helicity $\lambda _{\Omega}$
but this dependence will be canceled when divided by the bispinor
$\bar{u}_{\Omega}^{\mu} ( \bm{p}^{\prime} , \, \lambda _{\Omega} )
v_{\bar{\Omega} \, \mu} ( - \bm{p}^{\prime} , \, \lambda _{\Omega} )$
as in Eq.~\eqref{eq:ImV_gen}.

\section{Meson--meson scattering amplitude}
\label{app:meson}

\begin{figure}[!t]
  \centering
  \Psfig{8.6cm}{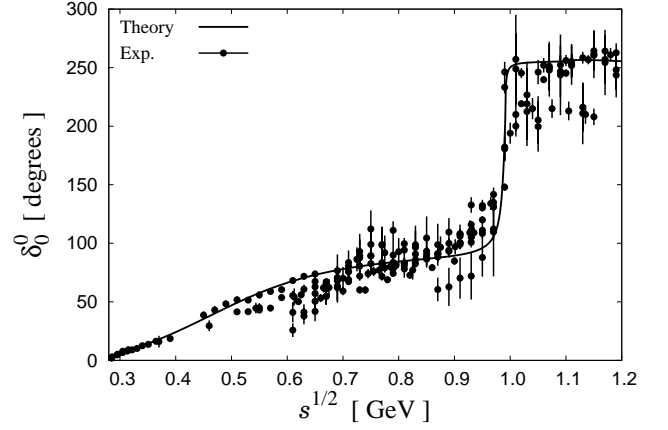}
  \caption{$\pi \pi ( J = 0 , \, I = 0 )$ phase shift $\delta
    _{0}^{0}$ in the present model.  Experimental data are taken from
    Refs.~\cite{Protopopescu:1973sh, Grayer:1974cr, Mukhin:1980,
      Kaminski:2001hv, Batley:2010zza}. }
  \label{fig:delta00}
\end{figure}

In this study we describe the $\pi \pi$-$K \bar{K}$-$\eta \eta$
coupled-channels scattering amplitude in the scalar-isoscalar channel
by using the so-called chiral unitary approach~\cite{Oller:1997ti,
  Oller:1997ng, Oller:1998hw, Oller:1998zr}.  In this approach we
calculate the scattering amplitude of two pseudoscalar mesons $T_{j
  k}^{\rm (2m)} ( s )$, where $j$ and $k$ are channel indices ($\pi
\pi$, $K \bar{K}$, and $\eta \eta$ for $j = 1$, $2$, and $3$,
respectively) and $s$ is the Mandelstam variable, by solving the
Lippmann--Schwinger equation in the following form:
\begin{equation}
  T_{j k}^{\rm (2m)} ( s ) = V_{j k}^{\rm (2m)} ( s )
  + \sum _{l} V_{j l}^{\rm (2m)} ( s ) G_{l}^{\rm (2m)} ( s )
  T_{l k}^{\rm (2m)} ( s ) ,
\end{equation}
with the interaction kernel $V_{j k} = V_{k j}$ taken from chiral
perturbation theory and the loop function of two pseudoscalar mesons
$G_{j}$.

We employ the leading-order terms of chiral perturbation theory for
the interaction kernel $V_{j k}$:
\begin{equation}
  \begin{split}
    & V_{1 1}^{\rm (2m)} ( s ) = \frac{m_{\pi}^{2} - 2 s}{f_{\pi}^{2}} ,
    \quad
    V_{1 2}^{\rm (2m)} ( s ) = - \frac{\sqrt{3} s}{2 \sqrt{2} f_{\pi} f_{K}} ,
    \\
    & V_{1 3}^{\rm (2m)} ( s ) = \frac{m_{\pi}^{2}}{\sqrt{3} f_{\pi} f_{\eta}} ,
    \quad
    V_{2 2}^{\rm (2m)} ( s ) = - \frac{3 s}{4 f_{K}^{2}} ,
    \\
    & V_{2 3}^{\rm (2m)} ( s ) = \frac{9 s - 2 m_{\pi}^{2} - 6 m_{\eta}^{2}}
      {6 \sqrt{2} f_{K} f_{\pi}} ,
    \\
    & V_{3 3}^{\rm (2m)} ( s ) = \frac{7 m_{\pi}^{2} - 16 m_{K}^{2}}{9 f_{\eta}^{2}}
    ,
  \end{split}
\end{equation}
where the on-shell approximation was used.  The loop function $G_{j}$
is evaluated with a three-dimensional sharp cutoff $q_{\rm max}$:
\begin{align}
  G_{j}^{\rm (2m)} ( s ) = & i N_{j} \int \frac{d^{4} q}{( 2 \pi )^{4}}
  \frac{1}{( q^{2} - m_{j}^{2} ) [ ( P - q )^{2} - m_{j}^{2} ]}
  \notag \\ = &
  \frac{N_{j}}{2 \pi ^{2}} \int _{0}^{q_{\rm max}} d q
  \frac{q^{2}}{\omega _{j} ( q )
    \left [ s - 4 \omega _{j} ( q )^{2} \right ]} ,
\end{align}
where $m_{j}$ is the meson mass in $j$th channel, $P^{\mu} = (
\sqrt{s} , \, \bm{0})$, $\omega _{j} ( q ) \equiv \sqrt{q^{2} +
  m_{j}^{2}}$, and the symmetry factor is $N_{1} = N_{3} = 1/2$ and
$N_{2} = 1$.  In this construction only the cutoff $q_{\rm max}$ is
the model parameter, and we fix it by fitting the $\pi \pi ( J = 0 ,
\, I = 0 )$ scattering phase shift $\delta _{0}^{0}$ to the
experimental data~\cite{Protopopescu:1973sh, Grayer:1974cr,
  Mukhin:1980, Kaminski:2001hv, Batley:2010zza} up to $\sqrt{s} = 1.2
\gev$.  From the best fit, we use the value of the cutoff $q_{\rm max}
= 850 \mev$, with which we can reproduce the phase shift $\delta
_{0}^{0}$ fairly well as shown in Fig.~\ref{fig:delta00}.


\begin{thebibliography}{99}

\bibitem{Weinberg:1965zz} 
  S.~Weinberg,
  Phys.\ Rev.\  {\bf 137}, B672 (1965).
  
\bibitem{Dyson:1964xwa} 
  F.~Dyson and N.~H.~Xuong,
  Phys.\ Rev.\ Lett.\  {\bf 13}, 815 (1964).

\bibitem{Jaffe:1976yi} 
  R.~L.~Jaffe,
  Phys.\ Rev.\ Lett.\  {\bf 38}, 195 (1977);
  \textit{ibid} {\bf 38}, 617 (1977).

\bibitem{Clement:2016vnl} 
  H.~Clement,
  Prog.\ Part.\ Nucl.\ Phys.\  {\bf 93}, 195 (2017).
  
\bibitem{Yamazaki:2002uh} 
  T.~Yamazaki and Y.~Akaishi,
  Phys.\ Lett.\ B {\bf 535}, 70 (2002).
  
\bibitem{Akaishi:2002bg} 
  Y.~Akaishi and T.~Yamazaki,
  Phys.\ Rev.\ C {\bf 65}, 044005 (2002).
  
\bibitem{Hyodo:2011ur} 
  T.~Hyodo and D.~Jido,
  Prog.\ Part.\ Nucl.\ Phys.\  {\bf 67}, 55 (2012).

\bibitem{Gal:2016boi} 
  A.~Gal, E.~V.~Hungerford and D.~J.~Millener,
  Rev.\ Mod.\ Phys.\  {\bf 88}, 035004 (2016).
    
\bibitem{Bashkanov:2008ih} 
  M.~Bashkanov {\it et al.},
  Phys.\ Rev.\ Lett.\  {\bf 102}, 052301 (2009).

\bibitem{Adlarson:2011bh} 
  P.~Adlarson {\it et al.} [WASA-at-COSY Collaboration],
  Phys.\ Rev.\ Lett.\  {\bf 106}, 242302 (2011).

\bibitem{Adlarson:2012fe} 
  P.~Adlarson {\it et al.} [WASA-at-COSY Collaboration],
  Phys.\ Lett.\ B {\bf 721}, 229 (2013).

\bibitem{Beane:2010hg} 
  S.~R.~Beane {\it et al.} [NPLQCD Collaboration],
  Phys.\ Rev.\ Lett.\  {\bf 106}, 162001 (2011).

\bibitem{Inoue:2010es} 
  T.~Inoue {\it et al.} [HAL QCD Collaboration],
  Phys.\ Rev.\ Lett.\  {\bf 106}, 162002 (2011).

\bibitem{Beane:2011iw} 
  S.~R.~Beane {\it et al.} [NPLQCD Collaboration],
  Phys.\ Rev.\ D {\bf 85}, 054511 (2012).

\bibitem{Inoue:2011ai} 
  T.~Inoue {\it et al.} [HAL QCD Collaboration],
  Nucl.\ Phys.\ A {\bf 881}, 28 (2012).

\bibitem{Sasaki:2013zwa} 
  K.~Sasaki {\it et al.} [HAL QCD Collaboration],
  Nucl.\ Phys.\ A {\bf 914}, 231 (2013).

\bibitem{Sasaki:2015ifa} 
  K.~Sasaki {\it et al.} [HAL QCD Collaboration],
  Prog.\ Theor.\ Exp.\ Phys.\ {\bf 2015}, 113B01 (2015).

\bibitem{Sasaki:2016gpc} 
  K.~Sasaki {\it et al.},
  PoS LATTICE {\bf 2015}, 088 (2016).
  
\bibitem{Sasaki:2017ysy} 
  K.~Sasaki {\it et al.},
  PoS LATTICE {\bf 2016}, 116 (2017).

\bibitem{Gongyo:2017fjb} 
  S.~Gongyo {\it et al.} [HAL QCD Collaboration],
  Phys.\ Rev.\ Lett., in press
  [arXiv:1709.00654 [hep-lat]].

\bibitem{Sada:2016nkb} 
  Y.~Sada {\it et al.} [J-PARC E15 Collaboration],
  Prog.\ Theor.\ Exp.\ Phys.\ {\bf 2016}, 051D01 (2016).

\bibitem{Sekihara:2016vyd} 
  T.~Sekihara, E.~Oset and A.~Ramos,
  Prog.\ Theor.\ Exp.\ Phys.\ {\bf 2016}, 123D03 (2016).

\bibitem{Goldman:1987ma} 
  J.~T.~Goldman, K.~Maltman, G.~J.~Stephenson~Jr., K.~E.~Schmidt and F.~Wang,
  Phys.\ Rev.\ Lett.\  {\bf 59}, 627 (1987).

\bibitem{Oka:1988yq} 
  M.~Oka,
  Phys.\ Rev.\ D {\bf 38}, 298 (1988).

\bibitem{Li:1999bc} 
  Q.~B.~Li and P.~N.~Shen,
  Eur.\ Phys.\ J.\ A {\bf 8}, 417 (2000).

\bibitem{Pang:2003ty} 
  H.~r.~Pang, J.~l.~Ping, F.~Wang, J.~T.~Goldman and E.~g.~Zhao,
  Phys.\ Rev.\ C {\bf 69}, 065207 (2004).

\bibitem{Zhu:2015sna} 
  X.~Zhu, H.~Huang, J.~Ping and F.~Wang,
  Phys.\ Rev.\ C {\bf 92}, 035210 (2015).

\bibitem{Huang:2015yza} 
  H.~Huang, J.~Ping and F.~Wang,
  Phys.\ Rev.\ C {\bf 92}, 065202 (2015).

\bibitem{Etminan:2014tya} 
  F.~Etminan {\it et al.} [HAL QCD Collaboration],
  Nucl.\ Phys.\ A {\bf 928}, 89 (2014).

\bibitem{Doi:2017zov} 
  T.~Doi {\it et al.},
  EPJ Web Conf.\  {\bf 175}, 05009 (2018).

\bibitem{Iritani:2018a} 
  T.~Iritani {\it et al.} [HAL QCD Collaboration],
  in preparation.  

\bibitem{Haidenbauer:2017sws} 
  J.~Haidenbauer, S.~Petschauer, N.~Kaiser, U.~G.~Mei{\ss}ner and W.~Weise,
  Eur.\ Phys.\ J.\ C {\bf 77}, 760 (2017).

\bibitem{Morita:2016auo} 
  K.~Morita, A.~Ohnishi, F.~Etminan and T.~Hatsuda,
  Phys.\ Rev.\ C {\bf 94}, 031901 (2016).

\bibitem{Olive:2016xmw} 
  C.~Patrignani [Particle Data Group Collaboration],
  Chin.\ Phys.\ C {\bf 40}, 100001 (2016).

\bibitem{Brown:1975di} 
  G.~E.~Brown and W.~Weise,
  Phys.\ Rept.\  {\bf 22}, 279 (1975).

\bibitem{Kim:1994ce} 
  H.~C.~Kim, J.~W.~Durso and K.~Holinde,
  Phys.\ Rev.\ C {\bf 49}, 2355 (1994).

\bibitem{Reuber:1995vc} 
  A.~Reuber, K.~Holinde, H.~C.~Kim and J.~Speth,
  Nucl.\ Phys.\ A {\bf 608}, 243 (1996).

\bibitem{Oller:1997ti} 
  J.~A.~Oller and E.~Oset,
  Nucl.\ Phys.\ A {\bf 620}, 438 (1997);
  \textit{ibid} {\bf 652}, 407 (1999).

\bibitem{Oller:1997ng} 
  J.~A.~Oller, E.~Oset and J.~R.~Pelaez,
  Phys.\ Rev.\ Lett.\  {\bf 80}, 3452 (1998).

\bibitem{Oller:1998hw} 
  J.~A.~Oller, E.~Oset and J.~R.~Pelaez,
  Phys.\ Rev.\ D {\bf 59}, 074001 (1999);
  \textit{ibid} {\bf 60}, 099906 (1999);
  \textit{ibid} {\bf 75}, 099903 (2007).

\bibitem{Oller:1998zr} 
  J.~A.~Oller and E.~Oset,
  Phys.\ Rev.\ D {\bf 60}, 074023 (1999).

\bibitem{Iritani:2018b} 
  T.~Iritani {\it et al.} [HAL QCD Collaboration],
  private communications.  

\bibitem{Ishikawa:2015rho} 
  K.-I.~Ishikawa {\it et al.} [PACS Collaboration],
  PoS LATTICE {\bf 2015}, 075 (2016).

\bibitem{Baru:2003qq} 
  V.~Baru, J.~Haidenbauer, C.~Hanhart, Y.~Kalashnikova and A.~E.~Kudryavtsev,
  Phys.\ Lett.\ B {\bf 586}, 53 (2004).

\bibitem{Sekihara:2014kya} 
  T.~Sekihara, T.~Hyodo and D.~Jido,
  Prog.\ Theor.\ Exp.\ Phys.\ {\bf 2015}, 063D04 (2015).

\bibitem{Sekihara:2014qxa} 
  T.~Sekihara and S.~Kumano,
  Phys.\ Rev.\ D {\bf 92}, 034010 (2015).

\bibitem{Sekihara:2016xnq} 
  T.~Sekihara,
  Phys.\ Rev.\ C {\bf 95}, 025206 (2017).

\bibitem{Hyodo:2011qc} 
  T.~Hyodo, D.~Jido and A.~Hosaka,
  Phys.\ Rev.\ C {\bf 85}, 015201 (2012).

\bibitem{Hyodo:2013nka} 
  T.~Hyodo,
  Int.\ J.\ Mod.\ Phys.\ A {\bf 28}, 1330045 (2013).

\bibitem{Formanek:2003}
  J. Form\'{a}nek, R. J. Lombard, J. Mare\v{s}
  Czech.\ J.\ Phys.\ {\bf 54}, 289 (2004).

\bibitem{Miyahara:2015bya} 
  K.~Miyahara and T.~Hyodo,
  Phys.\ Rev.\ C {\bf 93}, 015201 (2016).

\bibitem{Kamiya:2015aea} 
  Y.~Kamiya and T.~Hyodo,
  Phys.\ Rev.\ C {\bf 93}, 035203 (2016).

\bibitem{Kamiya:2016oao}%
  Y.~Kamiya and T.~Hyodo,
  Prog.\ Theor.\ Exp.\ Phys.\ \textbf{2017}, 023D02 (2017).

\bibitem{Cho:2010db} 
  S.~Cho {\it et al.} [ExHIC Collaboration],
  Phys.\ Rev.\ Lett.\  {\bf 106}, 212001 (2011).
  
\bibitem{Cho:2011ew} 
  S.~Cho {\it et al.} [ExHIC Collaboration],
  Phys.\ Rev.\ C {\bf 84}, 064910 (2011).

\bibitem{Cho:2017dcy} 
  S.~Cho {\it et al.} [ExHIC Collaboration],
  Prog.\ Part.\ Nucl.\ Phys.\  {\bf 95}, 279 (2017).

\bibitem{Jacob:1959at} 
  M.~Jacob and G.~C.~Wick,
  Annals Phys.\  {\bf 7}, 404 (1959);
  {\it ibid} {\bf 281}, 774 (2000).

\bibitem{Blankenbecler:1965gx} 
  R.~Blankenbecler and R.~Sugar,
  Phys.\ Rev.\  {\bf 142}, 1051 (1966).

\bibitem{Frazer:1960zza} 
  W.~R.~Frazer and J.~R.~Fulco,
  Phys.\ Rev.\  {\bf 117}, 1603 (1960).

\bibitem{Dumbrajs:1983jd} 
  O.~Dumbrajs, R.~Koch, H.~Pilkuhn, G.~c.~Oades, H.~Behrens, J.~j.~De Swart and P.~Kroll,
  Nucl.\ Phys.\ B {\bf 216}, 277 (1983).

\bibitem{Protopopescu:1973sh} 
  S.~D.~Protopopescu {\it et al.},
  Phys.\ Rev.\ D {\bf 7}, 1279 (1973).

\bibitem{Grayer:1974cr} 
  G.~Grayer {\it et al.},
  Nucl.\ Phys.\ B {\bf 75}, 189 (1974).

\bibitem{Mukhin:1980}
  K.~N.~Mukhin {\it et al.},
  JETP Lett.\  {\bf 32}, 601 (1980).

\bibitem{Kaminski:2001hv} 
  R.~Kaminski, L.~Lesniak and K.~Rybicki,
  Eur.\ Phys.\ J.\ direct C {\bf 4}, 4 (2002).

\bibitem{Batley:2010zza} 
  J.~R.~Batley {\it et al.} [NA48-2 Collaboration],
  Eur.\ Phys.\ J.\ C {\bf 70}, 635 (2010).

  
\end{thebibliography}
\end{document}